\documentclass[journal,10pt]{IEEEtran}

\makeatletter
\def\ps@headings{%
	\def\@oddhead{\mbox{}\scriptsize\rightmark \hfil \thepage}%
	\def\@evenhead{\scriptsize\thepage \hfil \leftmark\mbox{}}%
	\def\@oddfoot{}%
	\def\@evenfoot{}}
\makeatother 

\usepackage{graphicx}
\usepackage{amsmath, amsfonts,epsfig, multirow, floatflt,float}
\usepackage{amssymb}
\usepackage{cite}
\usepackage{algorithm,algorithmic}
\usepackage{hyperref}
\usepackage{enumitem}
\usepackage{mathrsfs}
\usepackage{url}
\usepackage{color}
\usepackage[caption=false,font=footnotesize,subrefformat=parens]{subfig}
\usepackage{array}
\usepackage{makecell}
\allowdisplaybreaks

\graphicspath{{figure/}}

\begin{document}
	
\title{Deep Reinforcement Learning Aided Platoon Control Relying on V2X Information}

\author{Lei~Lei {\it Senior Member, IEEE}, Tong~Liu, Kan~Zheng {\it Senior Member, IEEE}, Lajos Hanzo {\it Fellow, IEEE}
	
%
%
%
%
%
%
	
}

\maketitle

\begin{abstract}
The impact of Vehicle-to-Everything (V2X) communications on platoon control performance is investigated. Platoon control is essentially a sequential stochastic decision problem (SSDP), which can be solved by Deep Reinforcement Learning (DRL) to deal with both the control constraints and uncertainty in the platoon leading vehicle's behavior. In this context, the value of V2X communications for DRL-based platoon controllers is studied with an emphasis on the tradeoff between the gain of including exogenous information in the system state for reducing uncertainty and the performance erosion due to the curse-of-dimensionality. Our objective is to find the specific set of information that should be shared among the vehicles for the construction of the most appropriate state space. SSDP models are conceived for platoon control under different information topologies (IFT) by taking into account `just sufficient' information. Furthermore, theorems are established for comparing the performance of their optimal policies. In order to determine whether a piece of information should or should not be transmitted for improving the DRL-based control policy, we quantify its value by deriving the conditional KL divergence of the transition models. More meritorious information is given higher priority in transmission, since including it in the state space has a higher probability in offsetting the negative effect of having higher state dimensions. Finally, simulation results are provided to illustrate the theoretical analysis.       	  
\end{abstract}

	
	\section{Introduction}
     Autonomous vehicle platooning relies on a leading vehicle followed by a group of autonomous vehicles. The objective of platoon control is to determine the control input of the following autonomous vehicles so that all the vehicles move at the same speed while maintaining the desired distances between each pair of preceding and following vehicles. Platooning constitutes an efficient technique of increasing road capacity, reducing fuel consumption, as well as enhancing driving safety and comfort \cite{Li2017}. \par 
      
     Platoon control can be performed both with and without information exchange between vehicles using  Vehicle-to-Everything (V2X) communications. Platoon control without V2X is normally based on the adaptive cruise control (ACC) functionality, where the velocity of a following vehicle is autonomously adapted to keep a safe distance from its preceding vehicle based on its sensory information mainly obtained from radar. On the other hand, the more sophisticated cooperative adaptive cruise control (CACC) functionality extends ACC with V2X communication capabilities and it is capable of improving the platoon control performance by reducing the inter-vehicle distance of ACC \cite{Dey2016}.  \par

     \subsection{DRL-based Platoon Control}
     Platoon controllers have been proposed based on classical control theory, such as linear controller, $\mathcal{H}_{\infty}$ controller, and sliding mode controller (SMC) \cite{Li2017,Li2016}. On the other hand, platoon control is essentially a sequential stochastic decision problem (SSDP), where a sequence of decisions has to be made over a certain time horizon for a dynamic system whose state evolves in the face of uncertainty. The objective is to optimize the cumulative performance over the time horizon considered. The solution strategies to such a problem have been studied in different communities under different terminologies \cite{Powell2014}, such as stochastic optimal control/model predictive control (MPC) \cite{Todorov2005} in the control community, dynamic programming (DP)/approximate dynamic programming (ADP) \cite{Powell2011} in the Markov Decision Process (MDP) community, and reinforcement learning (RL)/deep reinforcement learning (DRL) \cite{Mnih2015} in the machine learning community.\par 
     
     The SSDP models conceived for platoon control generally have continuous state and action spaces. The optimal policies of such SSDP models can only be derived under the Linear-Quadratic-Guassian (LQG) formalism \cite{Todorov2005}, which is not the case for platoon control due to the uncertainty in the behavior of the leading vehicle and the state/control constraints. Therefore, techniques such as MPC and RL/DRL can be involved for deriving sub-optimal but practical policies. Various MPC strategies have been proposed in~\cite{Lan2020,Lin2020,MasseraFilho2017,vanNunen2019}, but there is a paucity of contributions relying on RL/DRL techniques~\cite{Desjardins2011,Wang2014,Huang2019,Zhang2020,Lin2019,Lin2021,Meixin2020,wei2018design,Buechel2018,Li2018,wang2018novel,Yan2021,Chu2019}.\par 
     
     To elaborate, the car-following control problem of supporting a single following vehicle has indeed been studied in a few contributions. In \cite{Desjardins2011}, DRL is used in a CACC system to learn high-level control policies on whether to brake, accelerate, or keep the current velocity. A learning proportional-integral (PI) controller is designed in~\cite{Wang2014}, where the parameters in the PI module are adaptively tuned based on the vehicle's state according to the control policy of the actor-critic learning module associated with kernel machines. However, a specific limitation of \cite{Wang2014} is that the candidate set of PI parameters has to be pre-determined. In order to avoid this problem, the parameterized batch actor-critic learning algorithm is proposed in~\cite{Huang2019} to generate the exact throttle/brake control input instead of the PI parameters. In~\cite{Zhang2020}, a deterministic RL method is conceived, which aims for improving the policy evaluation in the critic network and the exploration in the actor network. The acceleration-related delay is taken into account in \cite{Lin2019}, where a classical DRL algorithm - namely the Deep Deterministic Policy Gradient (DDPG) technique of \cite{lillicrap2019continuous} - is applied for an ACC system whose preceding vehicle is assumed to drive at a constant speed. The proposed algorithm is used for comparing the performance of DRL and MPC in~\cite{Lin2021}. In \cite{Meixin2020,wei2018design}, the human driving data has been used to help RL achieve improved performance. A velocity
       control scheme based on DDPG is proposed in~\cite{Meixin2020},
       where a reward function is developed by referencing human
       driving data and combining driving features related to safety,
       efficiency, and comfort. In~\cite{wei2018design}, a supervised
       RL-based framework is presented for the CACC system, where the
       actor and critic networks are updated under the guidance of the
       supervisor and the gain scheduler to improve the success rate
       of the training process. To learn a better control policy, the
       authors of~\cite{Buechel2018,Li2018} model/predict the
       leading vehicle's behavior. A predictive controller based on
       DDPG is presented in~\cite{Buechel2018}, which uses advance
       information about future speed reference values and road grade
       changes. A drift-mitigation oriented optimal control-based informed
       approximate Q-learning algorithm is developed for ACC systems
       in~\cite{Li2018}, where a hybrid Markov process is used to
       model the lead vehicle's speed. 
     
    For platoon control supporting multiple following
       vehicles, a CACC-based control algorithm using DDPG is proposed
       in \cite{wang2018novel}. In order to improve the platoon
       control performance, a hybrid strategy is advocated
       in~\cite{Yan2021} that selects the best actions obtained from
       the DDPG controller and a linear controller. In order to
       provide some safety guarantees to the control policy, a
       DDPG-based technique is invoked in~\cite{Chu2019} for
       determining the parameters of the optimal velocity model (OVM),
       which is in turn used to determine the vehicle accelerations.
     In contrast to most of the existing research relying on baseline
     DRL
     algorithms~\cite{Lin2019,Lin2021,Meixin2020,Buechel2018,wang2018novel,Chu2019},
     the Finite-Horizon DDPG (FH-DDPG) learning technique is adopted
     in this paper, which was proposed in our previous
     work~\cite{Lei2020} and proved to improve both the stability and the
     overall performance of the DDPG algorithm in a finite-horizon
     setting. Note that although the computational load of
       training a DRL agent is relatively high, the computational
       complexity for a trained DRL agent to make control decisions is
       very low during the deployment phase, since only the forward
       propagation in deep neural networks is involved. Moreover, the
       training of a DRL agent can be continued during the deployment
       phase in the background to keep improving control performance
       and adapt to new environment.  \par

     
     
     \subsection{Value of V2X Communications for Platoon Control}
     The beneficial impact of V2X communications on the performance of classical platoon controllers has been studied in \cite{Darbha1994,Konduri2017,Darbha2019,Zheng2016}. These platoon controllers are popularly designed by considering one of the following inter-vehicle spacing policies: Constant Spacing Policy (CSP) and Constant Time-headway Policy (CTHP) \cite{Li2017}. Explicitly, the desired distance between two adjacent vehicles is a constant value in CSP, while it is proportional to the vehicular speed in CTHP. As for CSP, it was demonstrated in \cite{Darbha1994} that a linear platoon controller purely relying on the information gleaned from the preceding vehicles but excluding the leading vehicle fails to guarantee string stability defined in \cite{Darbha1994}. This result is further verified in \cite{Konduri2017}. For CTHP, the benefits of using Vehicle-to-Vehicle (V2V) communications in terms of reducing the time headway required is investigated in \cite{Darbha2019}. In V2X communications, different information topologies (IFT) may be assumed, depending on the specific connectivity among the vehicles, such as the predecessor following (PF) type, the predecessor-leader following (PLF) type, and the bidirectional (BD) type \cite{Li2017}. In \cite{Zheng2016}, the influence of IFT on the internal stability and scalability of homogeneous vehicular platoons relying on linear feedback controllers was studied. \par
  
     However, there is a paucity of literature on quantifying the value of V2X communications for platoon controllers derived from solving SSDPs. In SSDP, the system evolves from one state to another as a result of decisions and exogenous information. A central challenge in solving SSDP is how to deal with one or more exogenous information processes, forcing us to make decisions before all the information becomes known \cite{Powell2011}. The extra information obtained through V2X might lead to the availability of sample realizations of the exogenous information before an action is determined, turning exogenous information into states in the SSDP models. This results in more informed platoon control decisions. \par
     
     However, exchanging large amount of V2X information incurs heavy communication overhead in vehicular networks \cite{Mei2018,Zhao2021}. Moreover, a complex state space may lead to another well-known challenge of dynamic programs, what is popularly termed as the curse-of-dimensionality. Planning in a reduced state space might in fact be more efficient than in the full model \cite{Chitnis2020}. DRL can be leveraged to alleviate the curse-of-dimensionality problem through function approximation by deep neural networks. However, the accuracy of the approximated value/policy functions might be reduced upon increasing the dimension of state space. To resolve this dilemma, our research addresses the research problem: \emph{what information should be transmitted between the vehicles through V2X communications to construct a sufficient yet compact state space for DRL-based platoon control}?  \par
	To the best of our knowledge, this paper is the first to analyze the value of V2X information for DRL-based platoon controllers. We boldly contrast our work to the existing works in Table \ref{table_contribution}. The contributions of this paper are itemized next.\par
      
     \begin{table*}[htb!]
     	\renewcommand{\arraystretch}{1}
     	\setlength{\extrarowheight}{1pt}
     	\centering
     	\caption{Summary of literature survey on platoon control}
     	\begin{tabular}{ |l|c|c|c|c|c|c|c|}
     		\hline
     		\multicolumn{2}{|c|}{}&\makecell[c]{\cite{Li2017,Li2016}\\ \cite{Lan2020,Lin2020,MasseraFilho2017,vanNunen2019}}&\makecell[c]{\cite{Wang2014,Huang2019,Zhang2020}\\\cite{wei2018design,Li2018}}&\makecell[c]{\cite{Desjardins2011,Buechel2018}\\\cite{Lin2019,Lin2021,Meixin2020}}&\cite{wang2018novel,Yan2021,Chu2019}&\cite{Darbha1994,Konduri2017,Darbha2019,Zheng2016}&Proposed \\
     		\cline{1-8}
     		\multicolumn{2}{|c|}{Classical platoon controller design} &\checkmark&&&&&\\
     		\hline
     		\multicolumn{2}{|c|}{RL-based platoon controller design}&&\checkmark&&&&\\ 
     		\cline{1-8}
     		\multirow{2}{*}{DRL-based platoon controller design}&Single following vehicle&&&\checkmark&&&\\ 
     		\cline{2-8}
     		&Multiple following vehicles&&&&\checkmark&& \\
     		\hline	
     		\multirow{2}{*}{Value for V2X information Analysis}&Classical platoon controller&&&&&\checkmark&\\ 
     		\cline{2-8}
     		&DRL-based platoon controller&&&&&&\checkmark \\
     		\hline
 	
     	\end{tabular}
     	\label{table_contribution}
     \end{table*}

     \subsection{Contributions} 
 
     \begin{itemize}
     	\item \textbf{A unified SSDP modeling framework}: While the RL/DRL theory is mostly developed based on the MDP model, the platoon control problems are more widely studied in the control community. In this paper, we define a general SSDP model unifying the terminologies from different communities, which may be conveniently used to formulate DRL-based platoon control problems.
     	\item \textbf{Value of V2X information for Optimal Platoon Control}: In order to address the question whether a piece of V2X information can be beneficially leveraged to improve the optimal policy of an SSDP problem, we formulate an augmented-state based SSDP when potentially useful V2X information becomes available, and provide theorems on when the optimal policy of an augmented-state problem could improve the original SSDP. With the aid of the proposed theorems, we are able to identify what V2X information and IFT are useful for improving the optimal control performance.  
     	\item \textbf{Value of V2X information for DRL-based Platoon Control}: Although the inclusion of V2X information in the state space promises to improve the performance of the optimal policy, larger state spaces might have a negative effect on the DRL-based policy performance due to its increased approximation errors in the value/policy functions. Therefore, even though a piece of V2X information has the potential to improve the optimal policy, it should not be transmitted and included in the state if it does not provide much meritorious information. In order to determine whether a piece of V2X information could help to improve the DRL-based policy, we quantify "\emph{How much better would we be able to predict the future state if we included the V2X information in the augmented-state}?" Specifically, we calculate the conditional KL divergence~\cite{Chitnis2020} of the probability distribution given by the product of the transition models of the original state and the V2X information, from the probability distribution given by the transition model of the augmented state. We then use it as a quantitative metric of characterizing the value of V2X information for DRL-based platoon control.
     \end{itemize} 

The remainder of the paper is organized as follows. The system model of platoon control is outlined in Section II. In Section III, we provide the definitions of both SSDP and augmented-state SSDP, and formulate general theorems for characterizing the value of exogenous information for SSDPs. Section IV uses the results of Section III to formulate the SSDP models of platoon control problems both with and without V2X communications. Then the performance of the optimal control policies of different SSDPs is compared. In Section V, the value of V2X information for DRL-based platoon control policies is evaluated based on the conditional KL divergence. Section VI reports on our simulations to validate the theoretical results. Finally, our conclusions are provided in Section VII.  

	\section{System Model for Platoon Control}
	\subsection{Two-Vehicle Scenario}
	  We first consider a simple vehicle-following control problem with only two vehicles, wherein the position, velocity and acceleration of a following vehicle (follower) $i$ at time $t$ are denoted by $p_{i}(t)$, $v_{i}(t)$, $acc_{i}(t)$, respectively. Here $p_{i}(t)$ represents the one-dimensional position of the center of the front bumper of vehicle $i$. \par
	
	The vehicle's dynamic model is described by
	\begin{equation}
	\label{eq2}
	\dot{p}_{i}(t)=v_{i}(t),
	\end{equation}
	\begin{equation}
	\label{eq3}
	\dot{v}_{i}(t)=acc_{i}(t),
	\end{equation}
	\begin{equation}
	\label{eq4}
	\dot{acc}_{i}(t)=-\frac{1}{\tau_{i}}acc_{i}(t)+\frac{1}{\tau_{i}}u_{i}(t),
	\end{equation}
	\noindent where $\tau_{i}=\tau$ is a time constant representing the driveline dynamics and $u_{i}(t)$ is the vehicle's control input at time instant $t$. In order to ensure safety and comfort, the following constraints are applied
	\begin{equation}
	\label{eq43}
	acc_{\mathrm{min}} \leq acc_{i}(t)\leq acc_{\mathrm{max}},
	\end{equation}
		\begin{equation}
	\label{eq44}
	u_{\mathrm{min}} \leq u_{i}(t)\leq u_{\mathrm{max}}.
	\end{equation}
	\noindent 
	 Note that \eqref{eq2}-\eqref{eq44} also apply to the preceding vehicle (predecessor) $i-1$ upon replacing the subscript $i$ by $i-1$. \par
	
	We denote the headway of vehicle $i$ at time $t$, i.e., bumper-to-bumper distance between $i$ and its predecessor $i-1$, by $d_{i}(t)$, which satisfies
	\begin{equation}
	\label{eq5}
	d_{i}(t)=p_{i-1}(t)-p_{i}(t)-L_{i-1},
	\end{equation}
	\noindent where $L_{i-1}$ is the the length of vehicle $i-1$.
	
	According to CTHP, vehicle $i$ aims for maintaining a desired headway of $d_{r,i}(t)$, given by
	\begin{equation}
	\label{eq6}
	d_{r,i}(t)=r_{i}+h_{i}v_{i}(t),
	\end{equation}
	\noindent where $r_{i}$ is a constant standstill distance for vehicle $i$ and $h_{i}$ is the desired time-gap of vehicle $i$.
	
	The control errors $e_{pi}(t)$ and $e_{vi}(t)$ are defined as
	\begin{equation}
	\label{eq7}
	e_{pi}(t)=d_{i}(t)-d_{r,i}(t),
	\end{equation}
	\begin{equation}
	\label{eq8}
	e_{vi}(t)=v_{i-1}(t)-v_{i}(t).
	\end{equation}

	Let $x_{i}(t)=[e_{pi}(t),e_{vi}(t),acc_{i}(t)]^{\mathrm{T}}$. The system dynamics evolve in continuous time according to
		\begin{equation}
		\label{eq30}
		\dot{x}_{i}(t)=A_{i}x_{i}(t)+B_{i}u_{i}(t)+C_{i}acc_{i-1}(t),
		\end{equation}
	\noindent where
	\begin{equation}
		\label{eq31}
A_{i}=\begin{bmatrix}
0 & 1 & -h_{i} \\
0 & 0 & -1 \\
0 & 0 & -\frac{1}{\tau_{i}}
\end{bmatrix},
B_{i}=\begin{bmatrix}
0\\
0 \\
\frac{1}{\tau_{i}}
\end{bmatrix},
C_{i}=\begin{bmatrix}
0\\
1 \\
0
\end{bmatrix}.		
\end{equation}

	\subsection{Platoon Scenario}
	We extend the two-vehicle scenario to a platoon that is composed of $N>2$ vehicles, i.e., $\mathcal{V}=\{0,1,\cdots,N-1\}$, where each vehicle $i\in\mathcal{V}$ obeys the dynamic model and the constraints given by \eqref{eq2}-\eqref{eq44}. Note that for the leading vehicle (leader) $0$, $e_{p0}(t)=e_{v0}(t)=0$, and we have
	\begin{equation}
	\label{eq15}
	\dot{x}_{0}(t)=A_{0}x_{0}(t)+B_{0}u_{0}(t),
	\end{equation}
	\noindent where
	\begin{equation}
	\label{eq16}
	A_{0}=\begin{bmatrix}
	0 & 0 & 0 \\
	0 & 0 & 0 \\
	0 & 0 & -\frac{1}{\tau_{0}}
	\end{bmatrix},
	B_{0}=\begin{bmatrix}
	0\\
	0 \\
	\frac{1}{\tau_{0}}
	\end{bmatrix}.	
	\end{equation}	
	For all the other vehicles $i\in\{1,2,\cdots,N-1\}$ in the
        platoon, the system dynamics evolve according to \eqref{eq30}
        and \eqref{eq31}. Note that the results in this
          paper may be applied to both homogeneous and heterogeneous
          platoons, where the vehicles can have the same or different
          dynamics. \par

	\subsection{System Dynamics in Discrete Time}
	In order to determine the vehicle's control action, an SSDP can be formulated. The time horizon is discretized into time intervals of length $T$ seconds (s), and a time period $[kT,(k+1)T)$ is referred to as a time step $k$, $k = 0,1,\cdots,K-1$, where $K$ is the total number of time steps. In the rest of the paper, we will use $x_{k}:=x(kT)$ to represent any variable $x$ at time $kT$. At each time step $k$, the controller of vehicle $i$ has to determine the vehicle's control action $u_{i,k}$. In this paper, we derive the system dynamics in discrete time based on forward Euler discretization of the dynamic system. \par


	\subsection{System State Observation by the Controller}
	The controller of vehicle $i$ has to determine $u_{i,k}$, $k=0,1,\cdots,K-1$ based on the observation of the system state at each time step. The velocity $v_{i,k}$ and acceleration $acc_{i,k}$ can be measured locally, while the control error $e_{pi,k}$ and $e_{vi,k}$ can be quantified by a radar unit mounted at the front of the vehicle. On the other hand, vehicle $i$ can only determine the driving status $x_{j,k}$ and vehicle control input $u_{j,k}$ of the other vehicles $j\in\mathcal{V}\backslash \{i\}$ through V2X communications.   \par
	
	In order to determine the optimal action $u_{i,k}$, $k=0,1,\cdots,K-1$, one salient question is, what information should be shared by V2X communications among vehicles, if any. We will answer this question by analyzing the value of exogenous information in an SSDP, where the related theory will be discussed in Section III. \par

	\section{Value of Exogenous Information in Sequential Stochastic Decision Problem}
	\subsection{SSDP Formulation}	
	\newtheorem{mydef}{Definition}
	\begin{mydef}[SSDP]
	Define an SSDP over a finite time horizon $k\in\{0,1,\ldots,K-1\}$ by $\{S_{k},a_{k},W_{k},f^{S},f^{W},R\}$, where $S_{k}\in\mathcal{S}$ and $a_{k}\in\mathcal{A}$ are the state and action in time step $k$ within state space $\mathcal{S}$ and action space $\mathcal{A}$\footnote{In the control community, the state and action are normally denoted by $x$ and $u$ (the latter is referred to as control) instead of $s$ and $a$. We adopt the current notation since it is more widely used in the RL/DRL community.}, respectively; $W_{k}\in\mathcal{W}$ is the exogenous information within its outcome space $\mathcal{W}$ that arrives during time step $k$ after decision $a_{k}$ has been made; $f^{S}$ is the system's state transition function governing $S_{k+1}=f^{S}(S_{k},a_{k},W_{k})$; $f^{W}$ is the transition function of the exogenous information $W_{k}$ governing $W_{k+1}=f^{W}(\{S_{k'}\}_{k'=0}^{k+1},\{a_{k'}\}_{k'=0}^{k+1},\{W_{k'}\}_{k'=0}^{k},\xi_{k})$, where $\xi_{k}$ represents all the parameters that affect the value of $W_{k+1}$ apart from the states and actions up to time step $k+1$, and exogenous information up to time step $k$. Furthermore, $R(S_{k},a_{k},W_{k})$ is the reward function. A policy $\pi=(\mu_{0},\ldots,\mu_{K-1})$ is a vector of functions $\mu_{k}$, where we have $a_{k}=\mu_{k}(S_{k})$ for each time step $k$. Under a policy $\pi$, the expected total reward $J_{\pi}$ over the finite time horizon can be expressed as
	\begin{equation}
	J_{\pi}=\max_{\pi}\{\mathrm{E}[\sum_{k=0}^{K-1}R(S_{k},\mu_{k}(S_{k}),W_{k})]\},
	\end{equation}
	The objective is to then find the optimal policy $\pi^{*}$ that maximizes the expected total reward, i.e.,
 	\begin{equation}
 \pi^{*}=\arg\max_{\pi}J_{\pi}.
 \end{equation}   
    \end{mydef}
   
  Note that in an SSDP as defined above, the decision $a_{k}$ is made in each time step $k$ solely based on state $S_{k}$ without knowing the exogenous information $W_{k}$. On the other hand, if the exogenous information $W_{k}$ is available at the time of making decisions for each time step, we can define an augmented-state SSDP.\par

     \newtheorem{mydef2}[mydef]{Definition}
  \begin{mydef2}[Augmented-state SSDP]
  	Assume that the exogenous information $W_{k}$ in the original SSDP given in Definition 1 is available before decision $a_{k}$ is made. Then we define an augmented-state SSDP by $\{\tilde{S}_{k},a_{k},\tilde{W}_{k},f^{\tilde{S}},f^{\tilde{W}},R\}$, where the augmented state $\tilde{S}_{k}=(S_{k},W_{k})$ is obtained by extending the state space of the original SSDP to include the additional information $W_{k}$. The action $a_{k}$ and the reward function $R(S_{k},a_{k},W_{k})=R(\tilde{S}_{k},a_{k})$ are the same as those of the original problem. The exogenous information is then given by
  	\begin{equation}
  	\label{eq1}
  	\tilde{W}_{k}=\{\{S_{k'}\}_{k'=0}^{k-1},\{a_{k'}\}_{k'=0}^{k-1},a_{k+1},\{W_{k'}\}_{k'=0}^{k-1},\xi_{k}\} 
  	\end{equation}
  	and the system's state transition function $f^{\tilde{S}}$ becomes  
  	\begin{align}
  	\label{eq10}
  	\tilde{S}_{k+1}&=
  	\begin{pmatrix}
  	S_{k+1}\\ W_{k+1}
  	\end{pmatrix} \IEEEnonumber \\	
  	& =   
  	\begin{pmatrix}
  	f^{S}(S_{k},a_{k},W_{k})\\ 
  	f^{W}(\{S_{k'}\}_{k'=0}^{k+1},\{a_{k'}\}_{k'=0}^{k+1},\{W_{k'}\}_{k'=0}^{k},\xi_{k})
  	\end{pmatrix} \IEEEnonumber \\	
  	&=f^{\tilde{S}}(\tilde{S}_{k},a_{k},\tilde{W}_{k}).
  	\end{align}
 Let us denote a policy as $\tilde{\pi}=(\tilde{\mu}_{0},\ldots,\tilde{\mu}_{K-1})$, where $a_{k}=\tilde{\mu}_{k}(\tilde{S}_{k})$ for each time step $k$. The transition function $f^{\tilde{W}}$ of exogenous information $\tilde{W}_{k}$ depends on $f^{S}$, $f^{W}$, $\tilde{\mu}_{k}$ and on the transition function for $\xi_{k}$.
  \end{mydef2} 
Note that the exogenous information given in \eqref{eq1}
  is derived from the third equality of \eqref{eq10} by comparing its
  L.H.S and R.H.S expressions. It can be seen that
  $\tilde{W}_{k}=\{S_{k},a_{k},W_{k}\}\cup\{\{S_{k'}\}_{k'=0}^{k+1},\{a_{k'}\}_{k'=0}^{k+1},\{W_{k'}\}_{k'=0}^{k},\xi_{k}\}\backslash\{\tilde{S}_{k},a_{k},S_{k+1}\}$. The
  reason that $S_{k+1}$ should be excluded from the exogenous
  information $\tilde{W}_{k}$ is due to the fact that given the
  augmented state $\tilde{S}_{k}$ and action $a_{k}$, $S_{k+1}$ can be
  determined by the transition function $f^{S}$.  
\newtheorem{remark}{Remark}
\begin{remark}[SSDP and MDP]
In Definition 1, we defined the SSDP, where the transition function of exogenous information $f^{W}$ considers the most general case. If we restrict the exogenous information transition function $f^{W}$ in Definition 1 to be $W_{k+1}=f^{W}(S_{k+1},a_{k+1},\xi_{k})$, where $\xi_{k}$ is an independent random variable with given distribution, the general SSDP reduces to an MDP.   
\end{remark}

	\subsection{Analyzing the Value of Exogenous Information}
\subsubsection{Value for the Optimal Policy}
Our objective is how to find out whether $\tilde{\pi}^{*}(\tilde{S}_{k})$ in Definition 2 will be improved over $\pi^{*}(S_{k})$ in Definition 1 as a result of exploiting $W_{k}$ before decision making. In the following, we provide three theorems that will be used in Section IV for analyzing the value of V2X communications for the optimal platoon control policies.  \par

  \newtheorem{thm}{Theorem}
\begin{thm}
	The optimal policy of the augmented-state SSDP $\tilde{\pi}^{*}(\tilde{S}_{k})$ is at least as good as that of the original SSDP $\pi^{*}(S_{k})$ if the exogenous information obeys $W_{k+1}=f^{W}(S_{k},W_{k},\xi_{k})$. Explicitly, $W_{k+1}$ depends on $S_{k}$ or $W_{k}$ or both, but not on other parameters except for $\xi_{k}$, which is an independent random variable. 
\end{thm}  
The proof of Theorem 1 is given in Appendix A. Physically, this suggests that the optimal policy could be improved, when the availability of exogenous information turns a non-Markovian SSDP into an MDP.\par

  \newtheorem{thm2}[thm]{Theorem}
\begin{thm2}
	The optimal policy of the augmented-state SSDP $\tilde{\pi}^{*}(\tilde{S}_{k})$ is at least as good as that of the original SSDP $\pi^{*}(S_{k})$ if the exogenous information obeys $W_{k+1}=f^{W}(S_{k+1},\xi_{k})$. Explicitly, $W_{k+1}$ may depend on $S_{k+1}$, but not on any other parameters except for $\xi_{k}$, which is an independent random variable. 
\end{thm2}  
The proof of Theorem 2 is given in Appendix B. We proved that the optimal policy may be improved by including the exogenous information in the state space even when the original SSDP is already an MDP.\par

  \newtheorem{thm3}[thm]{Theorem}
\begin{thm3}
	In Theorem 2, the optimal policy of the augmented-state SSDP $\tilde{\pi}^{*}(\tilde{S}_{k})$ has the same performance as that of the original SSDP $\pi^{*}(S_{k})$ if the exogenous information $W_{k}$ meets both of the following two conditions: (1) $W_{k}$ does not affect the transition of state $S_{k}$; and (2) $W_{k}$ does not affect the reward function. 
\end{thm3}  
The proof of Theorem 3 is given in Appendix C. Theorem 3 defines the conditions when the optimal policy cannot be improved by the availability of exogenous information.\par

\subsubsection{Value for the DRL-based Policy} Theorem 1 and 2 above provide the conditions when the availability of exogenous information $W_{k}$ can be leveraged for improving the optimal policy of an SSDP. However, having a larger state space may degrade the DRL-based policy's performance due to its reduced accuracy in the approximated value/policy functions. Therefore, we will quantify the value of $W_{k}$ and only include $W_{k}$ in the augmented-state $\tilde{S}_{k}$ when its value for improving the optimal policy is high enough to offset the negative effect of having a higher state dimension. In this way, we can construct the most appropriately dimensioned state space to derive DRL-based policies. \par

According to Theorem 3, the value of $W_{k}$ is related to the impact of $W_{k}$ on the transition of state $S_{k}$ and on the reward function. In the following, we will focus on the impact of $W_{k}$ on state transitions and propose a method of quantifying "\emph{How much better would we be able to predict the state $S_{k+1}$ if we included $W_{k}$ in the augmented-state $\tilde{S}_{k}$, versus we didn't}?" The proposed method will be used in Section V for evaluating the value of V2X information for DRL-based platoon control policies.  \par

Hence, we will first convert the transition functions for the system state and exogenous information, i.e., $f^{S}$, $f^{\tilde{S}}$, and $f^{W}$, to the corresponding transition probabilities $T^{S}=p\{S_{k+1}|S_{k},a_{k}\}$, $T^{\tilde{S}}=p\{S_{k+1},W_{k+1}|S_{k},a_{k},W_{k}\}$, and $T^{W}=p\{W_{k+1}|W_{k},S_{k},a_{k}\}$. Then, we will calculate the conditional KL divergence of $T^{S}\otimes T^{W}$ from $T^{\tilde{S}}$ as 
\begin{align}
\label{eq33}
&D_{KL}(T^{\tilde{S}}||T^{S}\otimes T^{W})=\int_{\tilde{S}_{k+1},\tilde{S}_{k},a_{k}}p\{\tilde{S}_{k+1},\tilde{S}_{k},a_{k}\} \IEEEnonumber \\
&\log\left(\frac{p\{\tilde{S}_{k+1}|\tilde{S}_{k},a_{k}\}}{p\{S_{k+1}|S_{k},a_{k}\}p\{W_{k+1}|S_{k},a_{k},W_{k}\}}\right). 
\end{align}
Note that the KL divergence in \eqref{eq33} is a measure of the information lost when $T^{S}\otimes T^{W}$ is used for approximating $T^{\tilde{S}}$. The KL divergence is $0$ if the transition of $S_{k}$ is independent of $W_{k}$. In this case, we know from Theorem 3 that the optimal policies of the augmented-state SSDP and the original SSDP are the same, and there is no need to include $W_{k}$ in $\tilde{S}_{k}$. On the other hand, a higher KL divergence value indicates that the transition of $S_{k}$ depends on $W_{k}$ to a larger extent, and thus the availability of $W_{k}$ is more important for accurately predicting the future state $S_{k+1}$. In this case, including $W_{k}$ in $\tilde{S}_{k}$ will be more likely to improve the DRL-based policy performance. Therefore, the KL divergence is a suitable quantitative measure for the value of the exogenous information.\par

\section{Value of V2X Communications For Optimal Platoon Control Policies}
	\subsection{SSDP for Two-Vehicle Scenario}
	In the following, we consider a two-vehicle scenario and assume that $u_{(i-1),k}$ of the predecessor is a sequence of independent random variables\footnote {We consider that the probability density function (pdf) of $u_{(i-1),k}$ is independent of the driving status and control input of vehicle $i$.}. We will formulate three SSDPs for the vehicle-following problem depending on whether V2X communications are available. Moreover, we will prove that better policies can be derived, when more information is available for the follower through V2X communications. \par
	
	In the rest of the paper, we will denote a policy to Problem $m$ by $\pi_{i}^{\mathrm{Pm}}=(\mu_{i,0}^{\mathrm{Pm}},\ldots,\mu_{i,K-1}^{\mathrm{Pm}})$. In Problem $m$, the objective is to find the optimal policy that maximizes the expected total reward, i.e., $\pi_{i}^{\mathrm{Pm}*}=\arg\max_{\pi_{i}^{\mathrm{Pm}}} J_{\pi_{i}^{\mathrm{Pm}}}$, and the expected total reward under the optimal policy is denoted by $J_{i}^{\mathrm{Pm}*}$. As the action space and reward functions of all the SSDPs are the same, we will only specify them in Problem 1.  \par
	
	\subsubsection{No V2X Communications}
	Without V2X communications, $acc_{(i-1),k}$ and $u_{(i-1),k}$ cannot be transmitted from the predecessor $i-1$ and become available for the follower $i$ to determine a vehicle control action $u_{i,k}$.

\newtheorem{problem}{Problem}
\begin{problem}[$\mathrm{P1}$]
The vehicle-following control problem operating without V2X communications can be formulated as an SSDP $\{S_{i,k}^{(\mathrm{P1})},a_{i,k},W_{i,k}^{(\mathrm{P1})},f^{S_{i}^{(\mathrm{P1})}},f^{W_{i}^{(\mathrm{P1})}},R\}$  with 
\begin{itemize}
	\item state $S_{i,k}^{(\mathrm{P1})}=x_{i,k}=[e_{pi,k},e_{vi,k},acc_{i,k}]^{\mathrm{T}}$;
	\item action $a_{i,k}=u_{i,k}$;
	\item exogenous information $W_{i,k}^{(\mathrm{P1})}=acc_{(i-1),k}$;
	\item system state transition function $f^{S_{i}^{\mathrm{P1}}}$ given by
	\begin{align}
	\label{eq11}
	S_{i,k+1}^{(\mathrm{P1})}=f^{S_{i}^{(\mathrm{P1})}}(S_{i,k}^{(\mathrm{P1})},a_{i,k},W_{i,k}^{(\mathrm{P1})}), 
	\end{align}	
	\noindent which can be derived from \eqref{eq30} and \eqref{eq31} based on forward Euler discretization;
	\item exogenous information transition function $f^{W_{i}^{(\mathrm{P1})}}$ given by
	\begin{align}
	\label{eq12}
	W_{i,k+1}^{(\mathrm{P1})}=(1-\frac{1}{\tau_{i}})W_{i,k}^{(\mathrm{P1})}+\frac{1}{\tau_{i}}u_{(i-1),k},
	\end{align}
	\noindent which can be derived from \eqref{eq4} based on forward Euler discretization;
	\item and the reward function $R(S_{i,k}^{(\mathrm{P1})},a_{i,k})$ given by 
	\begin{equation}
	\label{eq13}
	R(S_{i,k}^{(\mathrm{P1})},a_{i,k})=-\{(e_{pi,k})^2+\alpha(e_{vi,k})^2+\beta(a_{i,k})^2\},
	\end{equation}
	\noindent where $\alpha$ and $\beta$ are the weights
                   that are positive and can be adjusted
                    to determine the relative importance of minimizing
                    the position error, velocity error and the
                    control input.\par
\end{itemize}
\end{problem}

\subsubsection{With V2X communications}
With V2X communications, $acc_{(i-1),k}$ and $u_{(i-1),k}$ can be transmitted. In the following, we formulate two SSDPs depending on the transmitted information. 
\newtheorem{problem2}[problem]{Problem}
\begin{problem2}[$\mathrm{P2}$]
	The vehicle-following control problem relying on V2X communications where $acc_{(i-1),k}$ is transmitted from the preceding vehicle $i-1$ can be formulated as an SSDP $\{S_{i,k}^{(\mathrm{P2})},a_{i,k},W_{i,k}^{(\mathrm{P2})},f^{S_{i}^{(\mathrm{P2})}},f^{W_{i}^{(\mathrm{P2})}},R\}$ with 
	\begin{itemize}
		\item state $S_{i,k}^{(\mathrm{P2})}=[e_{pi,k},e_{vi,k},acc_{i,k},acc_{(i-1),k}]^{\mathrm{T}}=[(S_{i,k}^{(\mathrm{P1})})^{\mathrm{T}},W_{i,k}^{(\mathrm{P1})}]^{\mathrm{T}}$;
		\item exogenous information $W_{i,k}^{(\mathrm{P2})}=u_{(i-1),k}$, which is an independent random variable;
		\item system state transition function $f^{S_{i}^{(\mathrm{P2})}}$ given by
				\begin{align}
					\label{eq26}
		S_{i,k+1}^{(\mathrm{P2})} &= \begin{pmatrix}
		S_{i,k+1}^{(\mathrm{P1})}\\ W_{i,k+1}^{(\mathrm{P1})}
		\end{pmatrix}
		=
		\begin{pmatrix}
		f^{S_{i}^{(\mathrm{P1})}}(S_{i,k}^{(\mathrm{P1})},a_{i,k},W_{i,k}^{(\mathrm{P1})})\\ 
		f^{W_{i}^{(\mathrm{P1})}}(W_{i,k}^{(\mathrm{P1})},W_{i,k}^{(\mathrm{P2})}),
		\end{pmatrix}  \IEEEnonumber \\
		&=f^{S_{i}^{(\mathrm{P2})}}(S_{i,k}^{(\mathrm{P2})},a_{i,k},W_{i,k}^{(\mathrm{P2})})
		\end{align} 
		\noindent where $f^{S_{i}^{(\mathrm{P1})}}$ is formulated in \eqref{eq11}, and $f^{W_{i}^{(\mathrm{P1})}}$ is given in \eqref{eq12}.
		\item exogenous information  transition function $f^{W_{i}^{(\mathrm{P2})}}$ given by
		\begin{equation}
		\label{eq29}
		W_{i,k+1}^{(\mathrm{P2})}=f^{W_{i}^{(\mathrm{P2})}}(u_{i-1,k+1})=u_{i-1,k+1}.
		\end{equation}
	\end{itemize}
\end{problem2}	

\newtheorem{problem3}[problem]{Problem}
\begin{problem3}[$\mathrm{P3}$]
	The vehicle-following control problem harnessing V2X communications where $acc_{(i-1),k}$ and $u_{(i-1),k}$ are transmitted from the preceding vehicle $i-1$ can be formulated as an SSDP $\{S_{i,k}^{(\mathrm{P3})},a_{i,k},W_{i,k}^{(\mathrm{P3})},f^{S_{i}^{(\mathrm{P3})}},f^{W_{i}^{(\mathrm{P3})}},R\}$ with 
	\begin{itemize}
		\item state $S_{i,k}^{(\mathrm{P3})}=[e_{pi,k},e_{vi,k},acc_{i,k},acc_{(i-1),k},u_{(i-1),k}]^{\mathrm{T}}=[(S_{i,k}^{(\mathrm{P2})})^{\mathrm{T}},W_{i,k}^{(\mathrm{P2})}]^{\mathrm{T}}$;
		\item exogenous information $W_{i,k}^{(\mathrm{P3})}=u_{(i-1),(k+1)}$, i.e., the control input of the preceding vehicle $i-1$ in the next time step $k+1$, which is an independent random variable;
		\item system state transition function $f^{S_{i}^{(\mathrm{P3})}}$ given by
		\begin{align}
		S_{i,k+1}^{(\mathrm{P3})}&= \begin{pmatrix}
		S_{i,k+1}^{(\mathrm{P2})}\\ W_{i,k+1}^{(\mathrm{P2})}
		\end{pmatrix}
		=
		\begin{pmatrix}
		f^{S_{i}^{(\mathrm{P2})}}(S_{i,k}^{(\mathrm{P2})},a_{i,k},W_{i,k}^{(\mathrm{P2})})\\ 
		f^{W_{i}^{(\mathrm{P2})}}(W_{i,k}^{(\mathrm{P3})}),
		\end{pmatrix} \IEEEnonumber \\
		&=f^{S_{i}^{(\mathrm{P3})}}(S_{i,k}^{(\mathrm{P3})},a_{i,k},W_{i,k}^{(\mathrm{P3})})
		\end{align}
		\noindent where $f^{S_{i}^{(\mathrm{P2})}}$ is given in \eqref{eq26} and $f^{W_{i}^{(\mathrm{P2})}}$ is given in \eqref{eq29};
		\item exogenous information transition function $f^{W_{i}^{(\mathrm{P3})}}$ given by
		\begin{equation}
		W_{i,k+1}^{(\mathrm{P3})}=f^{W_{i}^{(\mathrm{P3})}}(u_{i-1,k+2})=u_{i-1,k+2}.
		\end{equation}
	\end{itemize}
\end{problem3}

\newtheorem{lem}{Lemma}
\begin{lem}
	\begin{enumerate}[label=\alph*.]
		\item The optimal policy $\pi_{i}^{\mathrm{P2}*}$ for SSDP $\mathrm{P2}$ performs at least as well as the optimal policy $\pi_{i}^{\mathrm{P1}*}$ for SSDP $\mathrm{P1}$, i.e., $J_{i}^{\mathrm{P2}*}\geq J_{i}^{\mathrm{P1}*}$. 
		\item The optimal policy $\pi_{i}^{\mathrm{P3}*}$ for SSDP $\mathrm{P3}$ performs at least as well as the optimal policy $\pi_{i}^{\mathrm{P2}*}$ for SSDP $\mathrm{P2}$, i.e., $J_{i}^{\mathrm{P3}*}\geq J_{i}^{\mathrm{P2}*}$.
	\end{enumerate}

\end{lem}

The proof of Lemma 1 is given in Appendix D. 

	\newtheorem{remark2}[remark]{Remark}
\begin{remark2}[Value of V2X information for the optimal vehicle-following policies]
Lemma 1a shows that transmission of the acceleration $acc_{(i-1),k}$ from the predecessor may result in improved optimal control performance of the follower $i$. Lemma 1b shows that the transmission of the control input $u_{(i-1),k}$ in addition to the acceleration from the predecessor can further improve the optimal control performance of follower $i$.\par 
\end{remark2}

	\subsection{SSDP for Platoon Control}
	We now consider the platooning scenario of $N>2$ vehicles and
        assume that $u_{0,k}$ of the leader $0$ is a sequence of
        independent random variables\footnote{We consider that the pdf
          of $u_{0,k}$ is independent of the driving status and
          control input of the following vehicles $i>0$.}. Consider
        that each vehicle $i>0$ determines its own control action
        $a_{i,k}$ in a decentralized fashion based on the state
        information received from its on-board sensors and V2X
        communications. Moreover, we focus on the scenario when the
        decentralized controls of the vehicles are coordinated, so
        that in each time step $k$, each vehicle $i$ makes control
        decisions only after all its predecessors\footnote{Note that a
          predecessor of vehicle $i$ refers to any vehicle in front of
          $i$ in the platoon, including the leader $0$.} $0\leq j<i$
        have made their control decisions $u_{j,k}$. The
          reason that we consider the above coordinated scenario is
          that in Lemma 1b, we have proved that the transmission of
          the control input $u_{(i-1),k}$ in addition to the
          acceleration from the predecessor can improve the optimal
          control performance of follower $i$. We consider the case
        when each vehicle $i>0$ only has to optimize its local reward
        $R(S_{i,k}^{(\mathrm{P1})},a_{i,k})$ defined in
        \eqref{eq13}. \par
	
	When no V2X communication is available, the decentralized platoon control problem reduces to SSDP $\mathrm{P1}$ for each vehicle $i>0$. In the following, we assume reliance on V2X communications.\par 
	
\subsubsection{V2X from the Immediate Predecessor $i-1$}
 When V2X communication is available to transmit $acc_{i-1,k}$ and $u_{i-1,k}$ from each immediate predecessor $i-1$ to its follower $i$, the decentralized platoon control problem reduces to an SSDP similar to $\mathrm{P3}$ for each vehicle $i>0$. However, an important difference between the decentralized platoon control problem and $\mathrm{P3}$ is that in the former, $u_{i-1,k}$ is no longer a sequence of independent random variables except for vehicle $i=1$.   \par
 
\newtheorem{problem4}[problem]{Problem}
\begin{problem4}[$\mathrm{P4}$]
	The decentralized platoon control problem for vehicle $i>0$ where $acc_{i-1,k}$ and $u_{i-1,k}$ are transmitted from its immediate predecessor $i-1$ can be formulated as an SSDP $\{S_{i,k}^{(\mathrm{P4})},a_{i,k},W_{i,k}^{(\mathrm{P4})},f^{S_{i}^{(\mathrm{P4})}},f^{W_{i}^{(\mathrm{P4})}},R\}$ with
	\begin{itemize}
		\item state $S_{i,k}^{(\mathrm{P4})}=[e_{pi,k},e_{vi,k},acc_{i,k},acc_{(i-1),k},u_{(i-1),k}]^{\mathrm{T}}=S_{i,k}^{(\mathrm{P3})}=[(S_{i,k}^{(\mathrm{P2})})^{\mathrm{T}},W_{i,k}^{(\mathrm{P2})}]^{\mathrm{T}}$;
		\item exogenous information $W_{i,k}^{(\mathrm{P4})}=\{u_{0,k+1}\}\cup W_{i,k}^{(\mathrm{P4\_1})}\cup W_{i,k}^{(\mathrm{P4\_2})}$, where $W_{i,k}^{(\mathrm{P4\_1})}=\{e_{p(i-1),k},e_{v(i-1),k}\}$ and $W_{i,k}^{(\mathrm{P4\_2})}=\{S_{j,k}^{(\mathrm{P1})}\}_{j=0}^{i-2}\cup \{u_{j,k}\}_{j=0}^{i-2}$;
		\item system state transition function $f^{S_{i}^{(\mathrm{P4})}}$ given by
		\begin{align}
		\label{eq32}
		S_{i,k+1}^{(\mathrm{P4})}&= \begin{pmatrix}
		S_{i,k+1}^{(\mathrm{P2})}\\ W_{i,k+1}^{(\mathrm{P2})}
		\end{pmatrix}
		=
		\begin{pmatrix}
		f^{S_{i}^{(\mathrm{P2})}}(S_{i,k}^{(\mathrm{P2})},a_{i,k},W_{i,k}^{(\mathrm{P2})})\\ 
		g^{W_{i}^{(\mathrm{P2})}}(S_{i,k}^{(\mathrm{P4})},W_{i,k}^{(\mathrm{P4})})
		\end{pmatrix} \IEEEnonumber \\
		&=f^{S_{i}^{(\mathrm{P4})}}(S_{i,k}^{(\mathrm{P4})},a_{i,k},W_{i,k}^{(\mathrm{P4})}),
		\end{align}
	    \noindent where $f^{S_{i}^{(\mathrm{P2})}}$ is given in \eqref{eq26}, while $g^{W_{i}^{(\mathrm{P2})}}$ is given by
		\begin{align}
		\label{eq22}
		W_{i,k+1}^{(\mathrm{P2})} =&u_{i-1,k+1}=\mu_{i-1}^{(\mathrm{P4})}(S_{i-1,k+1}^{(\mathrm{P4})})\IEEEnonumber \\
		=&\mu_{i-1}^{(\mathrm{P4})}(S_{i-1,k+1}^{(\mathrm{P2})},W_{i-1,k+1}^{(\mathrm{P2})})
		\IEEEnonumber \\
		=& \mu_{i-1}^{(\mathrm{P4})}(f^{S_{i-1}^{(\mathrm{P2})}}(S_{i-1,k}^{(\mathrm{P2})},u_{i-1,k},W_{i-1,k}^{(\mathrm{P2})}),W_{i-1,k+1}^{(\mathrm{P2})})
		\IEEEnonumber \\
		=&g^{W_{i}^{(\mathrm{P2})}}(\{S_{j,k}^{(\mathrm{P2})},u_{j,k},W_{j,k}^{(\mathrm{P2})}\}_{j=1}^{i-1},W_{1,k+1}^{(\mathrm{P2})}) \IEEEnonumber \\
		=& g^{W_{i}^{(\mathrm{P2})}}(\{S_{j,k}^{(\mathrm{P1})},u_{j,k}\}_{j=0}^{i-1},u_{0,k+1}) \IEEEnonumber \\
		=&g^{W_{i}^{(\mathrm{P2})}}((acc_{i-1,k},u_{i-1,k}),W_{i,k}^{(\mathrm{P4})}) \IEEEnonumber \\
		=&g^{W_{i}^{(\mathrm{P2})}}(S_{i,k}^{(\mathrm{P4})},W_{i,k}^{(\mathrm{P4})})
		\end{align}
		\noindent where	the fourth equality is derived upon iteratively replacing $W_{j,k+1}^{(\mathrm{P2})}$ by $\mu_{j-1}^{(\mathrm{P4})}(f^{S_{j-1}^{(\mathrm{P2})}}(S_{j-1,k}^{(\mathrm{P2})},u_{j-1,k},W_{j-1,k}^{(\mathrm{P2})}),W_{j-1,k+1}^{(\mathrm{P2})})$ for $j=\{i-1,\cdots,2\}$. Note that $W_{1,k+1}^{(\mathrm{P2})}=u_{0,k+1}$.
		\item exogenous information transition function $f^{W_{i}^{(\mathrm{P4})}}$ is given by
		\begin{align}
\label{eq41}
W_{i,k+1}^{(\mathrm{P4})}&= \begin{pmatrix}
u_{0,k+2}\\ W_{i,k+1}^{(\mathrm{P4\_1})} \\ W_{i,k+1}^{(\mathrm{P4\_2})}
\end{pmatrix} =
\begin{pmatrix}
u_{0,k+2}\\ 
f^{W_{i}^{(\mathrm{P4}\_1)}}(S_{i,k}^{(\mathrm{P4})},W_{i,k}^{(\mathrm{P4\_12})}) \\
f^{W_{i}^{(\mathrm{P4}\_2)}}(W_{i,k}^{(\mathrm{P4}\_2)}) \\
\end{pmatrix} \IEEEnonumber \\	
&=f^{W_{i}^{(\mathrm{P4})}}(W_{i,k}^{(\mathrm{P4})},S_{i,k}^{(\mathrm{P4})},u_{0,k+2}),
\end{align}
		\noindent where $W_{i,k}^{(\mathrm{P4}\_12)}=\{W_{i,k}^{(\mathrm{P4}\_1)},W_{i,k}^{(\mathrm{P4}\_2)}\}$. $f^{W_{i}^{(\mathrm{P4}\_1)}}(e_{p(i-1),k},e_{v(i-1),k},acc_{i-1,k},acc_{i-2,k})$ can be derived upon replacing $i$ by $i-1$ in $f^{S_{i}^{(\mathrm{P1})}}$. Note that $acc_{i-1,k}\in S_{i,k}^{(\mathrm{P4})}$, $\{e_{p(i-1),k},e_{v(i-1),k}\}\subset W_{i,k}^{(\mathrm{P4\_1})}$, and $acc_{i-2,k}\in W_{i,k}^{(\mathrm{P4\_2})}$. On the other hand, $f^{W_{i}^{(\mathrm{P4}\_2)}}=\{f^{S_{j}^{(\mathrm{P1})}},g^{W_{j+1}^{(\mathrm{P2})}}\}_{j=0}^{i-2}$, where $f^{S_{j}^{(\mathrm{P1})}}(S_{j,k}^{(\mathrm{P1})},u_{j,k},W_{j,k}^{(\mathrm{P1})})$ is given in \eqref{eq11}, while $g^{W_{j+1}^{(\mathrm{P2})}}(\{S_{j',k}^{(\mathrm{P1})},u_{j',k}\}_{j'=0}^{j},u_{0,k+1})$ is given in \eqref{eq22}.
	\end{itemize}
\end{problem4}

\begin{figure}[htb!]
	\centering
	\includegraphics[width=0.48\textwidth]{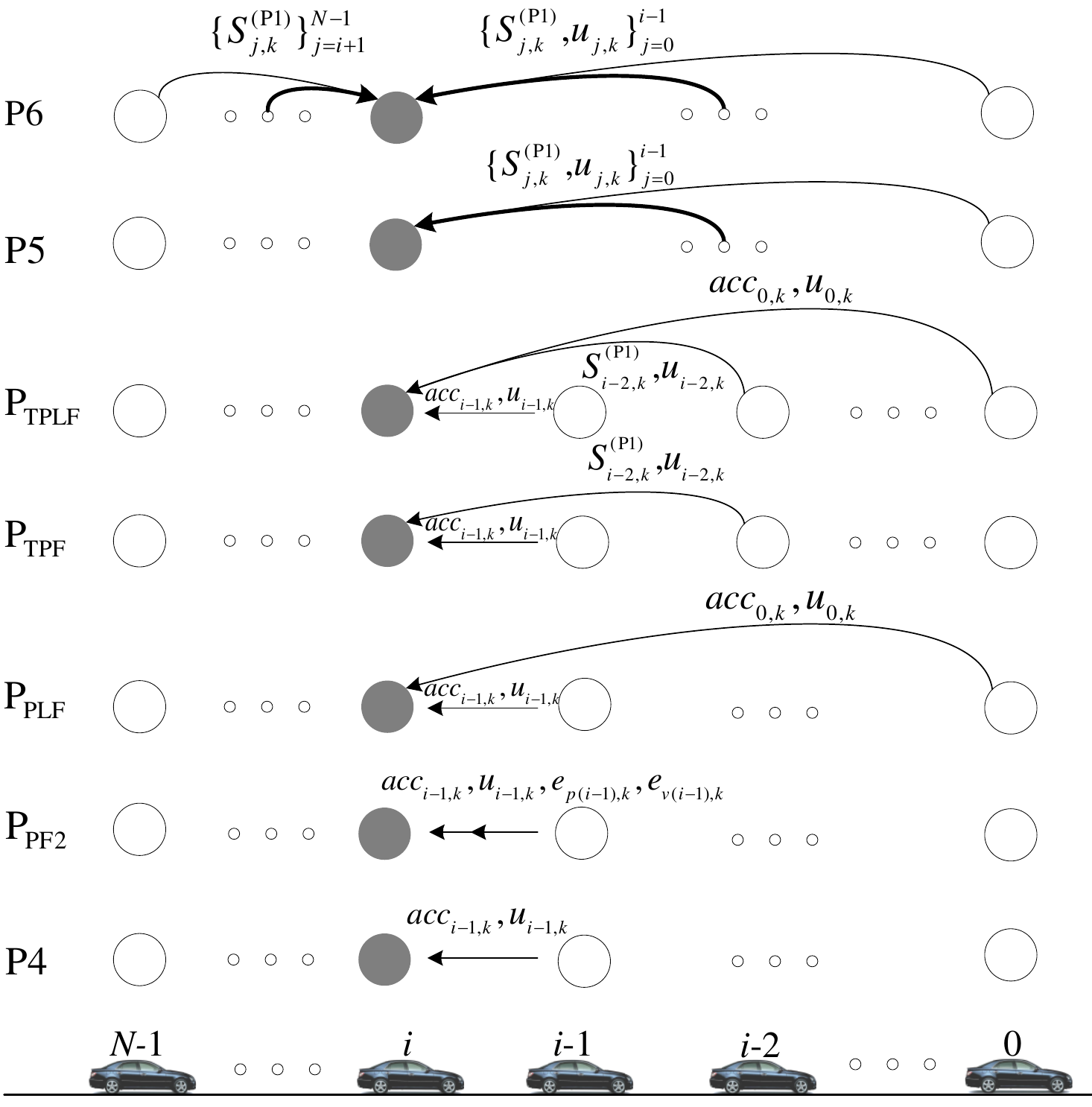}
	\caption{V2X information transmitted to vehicle $i$ in different SSDPs for a Platooning Scenario.}
	\label{fig_1}
\end{figure}

Fig.\ref{fig_1} shows the V2X information transmitted to vehicle $i$ in P4, while omitting those to other vehicles for a clear illustration.\par

\subsubsection{V2X from all Predecessors $\{0,\cdots,i-1\}$}
In SSDP $\mathrm{P4}$, since only $acc_{i-1,k}$ and $u_{i-1,k}$ are transmitted to vehicle $i$, the exogenous information $W_{i,k}^{(\mathrm{P4})}$ is not available for making control decisions. Note that although $u_{0,k+1}$ cannot be available at time step $k$, the rest of the information in $W_{i,k}^{(\mathrm{P4})}$, i.e., $W_{i,k}^{(\mathrm{P4}\_12)}$ can be made available to vehicle $i$. Note that in this case, the set of information $\{S_{j,k}^{(\mathrm{P1})},u_{j,k}\}_{j=0}^{i-1}$ has to be transmitted to vehicle $i$ from all its predecessors $0,\ldots,i-1$.  \par

\newtheorem{problem5}[problem]{Problem}
\begin{problem5}[$\mathrm{P5}$]
	The decentralized platoon control problem for vehicle $i>0$ where the information $\{S_{j,k}^{(\mathrm{P1})}\}_{j=0}^{i-1}\cup\{u_{j,k}\}_{j=0}^{i-1}$ is transmitted from all its predecessors $0,\ldots,i-1$ can be formulated as an SSDP $\{S_{i,k}^{(\mathrm{P5})},a_{i,k},W_{i,k}^{(\mathrm{P5})},f^{S_{i}^{(\mathrm{P5})}},f^{W_{i}^{(\mathrm{P5})}},R\}$ associated with 
	\begin{itemize}
		\item state $S_{i,k}^{(\mathrm{P5})}=[S_{0,k}^{(\mathrm{P1})},\cdots,S_{i,k}^{(\mathrm{P1})},u_{0,k},\cdots,u_{i-1,k}]=[(S_{i,k}^{(\mathrm{P4})})^{\mathrm{T}},(W_{i,k}^{(\mathrm{P4}\_12)})^{\mathrm{T}}]^{\mathrm{T}}$,
		\item exogenous information $W_{i,k}^{(\mathrm{P5})}=u_{0,k+1}$, which is a random variable with given distribution,
		\item system state transition function $f^{S_{i}^{(\mathrm{P5})}}$ given by
		\begin{align}
		\label{eq42}
		S_{i,k+1}^{(\mathrm{P5})}&= \begin{pmatrix}
		S_{i,k+1}^{(\mathrm{P4})}\\ W_{i,k+1}^{(\mathrm{P4}\_1)}\\W_{i,k+1}^{(\mathrm{P4}\_2)}
		\end{pmatrix}
		=
		\begin{pmatrix}
		f^{S_{i}^{(\mathrm{P4})}}(S_{i,k}^{(\mathrm{P4})},a_{i,k},W_{i,k}^{(\mathrm{P4})})\\ 
		f^{W_{i}^{(\mathrm{P4}\_1)}}(S_{i,k}^{(\mathrm{P4})},W_{i,k}^{(\mathrm{P4\_12})}) \\
		f^{W_{i}^{(\mathrm{P4}\_2)}}(W_{i,k}^{(\mathrm{P4}\_2)}),
		\end{pmatrix} \IEEEnonumber \\
		&=f^{S_{i}^{(\mathrm{P5})}}(S_{i,k}^{(\mathrm{P5})},a_{i,k},W_{i,k}^{(\mathrm{P5})}),
		\end{align}
        \noindent where $f^{S_{i}^{(\mathrm{P4})}}$ is given in \eqref{eq32} and $f^{W_{i}^{(\mathrm{P4}\_1)}}$ and $f^{W_{i}^{(\mathrm{P4}\_2)}}$ are given in \eqref{eq41}.
		\end{itemize}
\end{problem5}	

Fig.\ref{fig_1} shows the V2X information transmitted to vehicle $i$ in $\mathrm{P5}$ and omitted those to other vehicles for a clear illustration.\par

\newtheorem{lem2}[lem]{Lemma}
\begin{lem2}
	The optimal policy $\pi_{i}^{\mathrm{P5}*}$ for SSDP $\mathrm{P5}$ performs at least as well as the optimal policy $\pi_{i}^{\mathrm{P4}*}$ for SSDP $\mathrm{P4}$, i.e., we have $J_{i}^{\mathrm{P5}*}\geq J_{i}^{\mathrm{P4}*}$.
\end{lem2}

The proof of Lemma 2 is given in Appendix E. In $\mathrm{P5}$, note that we assume that the control policies for all the predecessors $1,\cdots,i-1$ are derived from $\mathrm{P4}$. However, it can be proved that when the control policies of all the predecessors  $1,\cdots,i-1$ are derived from $\mathrm{P5}$, Lemma 2 is still valid. Due to space limitation, we will omit the detailed proof in this paper.\par


\subsubsection{V2X from all Other Vehicles $\{0,\cdots,N-1\}\backslash\{i\}$}
\newtheorem{problem6}[problem]{Problem}
\begin{problem6}[$\mathrm{P6}$]
	The decentralized platoon control problem for vehicle $i>0$ where the information $\{S_{j,k}^{(\mathrm{P1})}\}_{j=0}^{i-1}\cup\{u_{j,k}\}_{j=0}^{i-1}$ is transmitted from all its predecessors $0,\ldots,i-1$ and the information $\{S_{j,k}^{(\mathrm{P1})}\}_{j=i+1}^{N-1}$ is transmitted from all of its followers $i+1,\ldots,N-1$ can be formulated as an SSDP $\{S_{i,k}^{(\mathrm{P6})},a_{i,k},W_{i,k}^{(\mathrm{P6})},f^{S_{i}^{(\mathrm{P6})}},f^{W_{i}^{(\mathrm{P6})}},R^{(\mathrm{P6})}\}$ with state $S_{i,k}^{(\mathrm{P6})}=[S_{0,k}^{(\mathrm{P1})},\cdots,S_{N-1,k}^{(\mathrm{P1})},u_{0,k},\cdots,u_{i-1,k}]^{\mathrm{T}}=[(S_{i,k}^{(\mathrm{P5})})^{\mathrm{T}},(\bar{W}_{i,k}^{(\mathrm{P5})})^{\mathrm{T}}]^{\mathrm{T}}$, where $\bar{W}_{i,k}^{(\mathrm{P5})}=[S_{i+1,k}^{(\mathrm{P1})},\cdots,S_{N-1,k}^{(\mathrm{P1})}]^{\mathrm{T}}$.
\end{problem6}	

Fig.\ref{fig_1} shows the V2X information transmitted to vehicle $i$ in $\mathrm{P6}$, where we omit those to other vehicles for a clear illustration.\par

\newtheorem{lem3}[lem]{Lemma}
\begin{lem3}
	The optimal policy $\pi_{i}^{\mathrm{P6}*}$ for SSDP $\mathrm{P6}$ has the same performance as the optimal policy $\pi_{i}^{\mathrm{P5}*}$ for SSDP $\mathrm{P5}$, i.e., we have $J_{i}^{\mathrm{P6}*}=J_{i}^{\mathrm{P5}*}$.
\end{lem3}
The proof of Lemma 3 is given in Appendix F.

	\newtheorem{remark3}[remark]{Remark}
\begin{remark3}[Value of V2X information for the optimal platoon control policies]
Lemma 2 shows that the transmission of the driving status $S_{j,k}^{(\mathrm{P1})}$ and control input $u_{j,k}$ from all the predecessors $0\leq j<i$ instead of only the acceleration $acc_{(i-1),k}$ and control input $u_{(i-1),k}$ of the immediate predecessor $i-1$ to vehicle $i$ may improve the optimal control performance. Lemma 3 shows that the transmission of the driving status $S_{j,k}^{(\mathrm{P1})}$ from the followers $i+1<j\leq N-1$ of vehicle $i$ cannot help vehicle $i$ improving the optimal control decisions. 
\end{remark3}

\section{Value of V2X Communications for DRL-based Platoon Control Policies}
\subsection{Value of $W_{i,k}^{(\mathrm{P4}\_12)}$}
From Remark 3, we can see that the transmission of information from all the predecessors of vehicle $i>0$ in the platoon instead of only its immediate predecessor may improve its optimal control policy. In other words, with the transmission of additional information $W_{i,k}^{(\mathrm{P4}\_12)}$, $\mathrm{P5}$ can be formulated with an optimal policy performing at least as well as $\mathrm{P4}$. However, $\mathrm{P5}$ involves higher communication and computation overheads than $\mathrm{P4}$. As discussed in Section III.B-2), we should only include substantial exogenous information for predicting future states in the augmented-state to get improved DRL-based control policy. Therefore, we will quantify the value of $W_{i,k}^{(\mathrm{P4}\_12)}$ according to the method proposed in Section III.B-2).\par

Firstly, we convert the transition functions of $\mathrm{P4}$ and $\mathrm{P5}$ to transition probabilities as below. Specifically, the system state transition probability for $\mathrm{P5}$ is
\begin{align}
\label{eq34}
&T^{S_{i}^{(\mathrm{P5})}}=p\{S_{i,k+1}^{(\mathrm{P5})}|S_{i,k}^{(\mathrm{P5})},a_{k}\} \IEEEnonumber \\
=&\prod_{j=0}^{i}\mathbf{1}_{S_{j,k+1}^{(\mathrm{P1})}=f^{S_{j}^{(\mathrm{P1})}}(S_{j,k}^{(\mathrm{P1})},u_{j,k},acc_{j-1,k})}\prod_{j=1}^{i-1}p\{u_{j,k+1}|\{S_{j',k}^{(\mathrm{P1})},\IEEEnonumber \\&u_{j',k}\}_{j'=0}^{j}\}p\{u_{0,k+1}\},
\end{align} 
\noindent where $\mathbf{1}_{X}$ is $1$ when $X$ is true and $0$ otherwise.\par
The system's state transition probability for $\mathrm{P4}$ is
\begin{align}
\label{eq35}
&T^{S_{i}^{(\mathrm{P4})}}=p\{S_{i,k+1}^{(\mathrm{P4})}|S_{i,k}^{(\mathrm{P4})},a_{k}\}
\IEEEnonumber \\
=&\mathbf{1}_{S_{i,k+1}^{(\mathrm{P1})}=f^{S_{i}^{(\mathrm{P1})}}(S_{i,k}^{(\mathrm{P1})},u_{i,k},acc_{i-1,k})}p\{u_{i-1,k+1}|acc_{i-1,k},u_{i-1,k}\}\IEEEnonumber \\
&\mathbf{1}_{acc_{i-1,k+1}=f^{W_{i}^{(\mathrm{P1})}}(acc_{i-1,k},u_{i-1,k})}.
\end{align} 
The exogenous information ($W_{i,k}^{(\mathrm{P4}\_12)}$) transition probability for $\mathrm{P4}$ is
\begin{align}
\label{eq36}
&T^{W_{i}^{(\mathrm{P4\_12})}}=p\{W_{i,k+1}^{(\mathrm{P4\_12})}|S_{i,k}^{(\mathrm{P4})},a_{k},W_{i,k}^{(\mathrm{P4\_12})}\}
\IEEEnonumber \\=&\prod_{j=0}^{i-2}\mathbf{1}_{S_{j,k+1}^{(\mathrm{P1})}=f^{S_{j}^{(\mathrm{P1})}}(S_{j,k}^{(\mathrm{P1})},u_{j,k},acc_{j-1,k})}\prod_{j=1}^{i-2}p\{u_{j,k+1}|\{S_{j',k}^{(\mathrm{P1})},\IEEEnonumber \\&u_{j',k}\}_{j'=0}^{j}\}p\{u_{0,k+1}\}
\mathbf{1}_{\substack{e_{p(i-1),k+1},e_{v(i-1),k+1}=f^{W_{i}^{(\mathrm{P4}\_1)}}(e_{p(i-1),k},\\e_{v(i-1),k},acc_{i-1,k},acc_{i-2,k})}}.
\end{align} 

Next, we derive the conditional KL divergence of $T^{S_{i}^{(\mathrm{P4})}}\otimes T^{W_{i}^{(\mathrm{P4\_12})}}$ from $T^{S_{i}^{(\mathrm{P5})}}$ as
\begin{align}
\label{eq37}
&D_{KL}(T^{S_{i}^{(\mathrm{P5})}}||T^{S_{i}^{(\mathrm{P4})}}\otimes T^{W_{i}^{(\mathrm{P4\_12})}})\IEEEnonumber \\=&\int_{S_{i,k+1}^{(\mathrm{P5})},S_{i,k}^{(\mathrm{P5})},a_{k}}p\{S_{i,k+1}^{(\mathrm{P5})},S_{i,k}^{(\mathrm{P5})},a_{k}\}\IEEEnonumber \\&\log\left(\frac{p\{S_{i,k+1}^{(\mathrm{P5})}|S_{i,k}^{(\mathrm{P5})},a_{k}\}}{p\{S_{i,k+1}^{(\mathrm{P4})}|S_{i,k}^{(\mathrm{P4})},a_{k}\}p\{W_{i,k+1}^{(\mathrm{P4\_12})}|S_{i,k}^{(\mathrm{P4})},W_{i,k}^{(\mathrm{P4\_12})}\}}\right)\IEEEnonumber \\
=&\int_{u_{i-1,k+1},\{S_{j,k}^{(\mathrm{P1})},u_{j,k}\}_{j=0}^{i-1}}p\big\{u_{i-1,k+1},\{S_{j,k}^{(\mathrm{P1})},u_{j,k}\}_{j=0}^{i-1}\big\}\IEEEnonumber \\
&\log\left(\frac{p\big\{u_{i-1,k+1}|\{S_{j,k}^{(\mathrm{P1})},u_{j,k}\}_{j=0}^{i-1}\big\}}{p\{u_{i-1,k+1}|acc_{i-1,k},u_{i-1,k}\}}\right).
\end{align}
From \eqref{eq37}, we can see that the KL divergence depends on the
ratio between
$p\{u_{i-1,k+1}|\{S_{j,k}^{(\mathrm{P1})},u_{j,k}\}_{j=0}^{i-1}\}$ and
$p\{u_{i-1,k+1}|acc_{i-1,k},u_{i-1,k}\}$, i.e., \emph{how much better
  we can predict the control input $u_{i-1,k+1}$ of the predecessor
  $i-1$ in the next time step, given the additional information
  $W_{i,k}^{(\mathrm{P4}\_12)}$}? Given the trained actor network for
vehicle $i-1$, the empirical value of this ratio can be obtained by
Monte Carlo simulation. For $e$ episodes of experiences
  obtained through Monte Carlo simulation, the computational
  complexity of calculating the KL divergence using \eqref{eq37} in
  $O(e^2)$.\par

\subsection{Value of Components in $W_{i,k}^{(\mathrm{P4}\_12)}$}
In the above analysis, we assumed that either all or none of the information in $W_{i,k}^{(\mathrm{P4}\_12)}$ is transmitted to vehicle $i$. However, we could strike a better tradeoff between improving the performance of the optimal policy and reducing the state space dimension by including only the components in $W_{i,k}^{(\mathrm{P4}\_12)}$ that have high value in helping to better predict the future state $S_{i,k+1}^{(\mathrm{P4})}$. In this way, we hope to reduce the communication overhead and improve the DRL-based policy. \par

Interestingly, the inclusion of different components in $W_{i,k}^{(\mathrm{P4}\_12)}$ can be aligned with the typical IFT for the platoon \cite{Zheng2016}. As we only focus on the specific IFT in which information was transmitted only from predecessors but not followers as discussed in Remark 3, we examine the following four typical IFTs listed below:
\begin{itemize}
	\item \textbf{PF topology}: Problem 4 is actually based on PF, where only the immediate predecessor $i-1$ transmits information to vehicle $i$. In Problem 4, only $acc_{i-1,k}$ and $u_{i-1,k}$ are transmitted. In addition, $W_{i,k}^{(\mathrm{P4}\_1)}=\{e_{p(i-1),k},e_{v(i-1),k}\}$ could also be transmitted.
	\item \textbf{PLF topology}: Not only the immediate predecessor $i-1$, but also the leader $0$ transmit information $S_{0,k}^{(\mathrm{P1})}$ (i.e., $acc_{0,k}$) and $u_{0,k}$ to vehicle $i$. 
	\item \textbf{Two-predecessors following (TPF) topology}: Not only the immediate predecessor $i-1$, but also the second immediate predecessor $i-2$ transmit information $S_{i-2,k}^{(\mathrm{P1})},u_{i-2,k}$ to vehicle $i$. 
	\item \textbf{Two-predecessor-leader following (TPLF) topology}: Not only the immediate predecessor $i-1$, but also the second immediate predecessor $i-2$ and the leader $0$ transmit information to vehicle $i$. 
\end{itemize} 

According to the different IFT and V2X information, we can formulate a number of SSDPs as seen in Table \ref{table_4} and illustrated in Fig.\ref{fig_1}. Note again that Fig.\ref{fig_1} only shows the V2X information transmitted to vehicle $i$  and omitted those to other vehicles for avoiding obfuscation. The state $S_{i,k}^{(\mathrm{Pm})}$ of any SSDP $\mathrm{Pm}$ in Table \ref{table_4} includes the driving status $S_{i,k}^{(\mathrm{P1})}$ of vehicle $i$ as well as the V2X information $I_{i,k}^{(\mathrm{Pm})}$ transmitted to vehicle $i$, i.e.,  $S_{i,k}^{(\mathrm{Pm})}=\{S_{i,k}^{(\mathrm{P1})},I_{i,k}^{(\mathrm{Pm})}\}$.  \par
\begin{table}
	\centering
	\caption{IFT and V2X information for different SSDPs in Platoon Scenario}
	\begin{tabular}{ llr }
		\hline
		\textbf{SSDP} & \textbf{IFT} & \textbf{V2X information} $I_{i,k}^{(\mathrm{Pm})}$ \\
		\hline
		$\mathrm{P4}$ & PF & $acc_{i-1,k}$, $u_{i-1,k}$ \\ 
		\hline
		$\mathrm{P_{PF2}}$ & PF & $acc_{i-1,k}$, $u_{i-1,k}$, $e_{p(i-1),k}$, $e_{v(i-1),k}$ \\ 
		\hline
		$\mathrm{P_{PLF}}$ & PLF & $acc_{i-1,k}$, $u_{i-1,k}$, $acc_{0,k}$, $u_{0,k}$  \\ 
		\hline
		$\mathrm{P_{TPF}}$ & TPF & $acc_{i-1,k}$, $u_{i-1,k}$, $S_{i-2,k}^{(\mathrm{P1})}$, $u_{i-2,k}$ \\ 
		\hline
		$\mathrm{P_{TPLF}}$ & TPLF & $acc_{i-1,k}$, $u_{i-1,k}$, $S_{i-2,k}^{(\mathrm{P1})}$, $u_{i-2,k}$, $acc_{0,k}$, $u_{0,k}$  \\
		\hline
		$\mathrm{P5}$ & - & $\{S_{j,k}^{(\mathrm{P1})},u_{j,k}\}_{j=0}^{i-1}$\\
		\hline
	\end{tabular}
	\label{table_4}
\end{table}

In Table \ref{table_4}, if the state $S_{i,k}^{(\mathrm{Pm})}$ of an SSDP $\mathrm{Pm}$ is a subset of the state $S_{i,k}^{(\mathrm{Pn})}$ of another SSDP $\mathrm{Pn}$ (e.g., the state of $\mathrm{P4}$ is a subset of all the other SSDPs in Table \ref{table_4}), we can analyze the value of additional information $S_{i,k}^{(\mathrm{Pn})}\backslash S_{i,k}^{(\mathrm{Pm})}$ by deriving the KL divergence for including the additional information as
	\begin{align}
	\label{eq38}
	&D_{KL}(T^{S_{i,k}^{(\mathrm{Pn})}}||T^{S_{i,k}^{(\mathrm{Pm})}}\otimes T^{S_{i,k}^{(\mathrm{Pn})}\backslash S_{i,k}^{(\mathrm{Pm})}})\IEEEnonumber \\
	=&\int_{\substack{u_{i-1,k+1}\\I_{i,k}^{(\mathrm{Pn})}}}p\{u_{i-1,k+1},I_{i,k}^{(\mathrm{Pn})}\}\log\left(\frac{p\{u_{i-1,k+1}|I_{i,k}^{(\mathrm{Pn})}\}}{p\{u_{i-1,k+1}|I_{i,k}^{(\mathrm{Pm})}\}}\right).
	\end{align} 
\noindent Note that \eqref{eq37} can be considered as a special case of \eqref{eq38} when $\mathrm{Pm=P4}$ and $\mathrm{Pn=P5}$. Similar to \eqref{eq37}, the KL divergence in \eqref{eq38} depends on \emph{how much better we can predict the control input $u_{i-1,k+1}$ of the predecessor $i-1$ in the next time step given the additional information $S_{i,k}^{(\mathrm{Pn})}\backslash S_{i,k}^{(\mathrm{Pm})}$}? \par

\section{Experimental Results}
In this section, we present our simulation results of the DRL-based platoon control policies for different IFT and V2X information. The platoon control environment and the DRL algorithms are implemented in Tensorflow 1.14 using Python.

\subsection{Experimental Setup}

The technical constraints and operational parameters of the platoon control environment are given in Table~\ref{table_1} \cite{Lin2019}. The interval for each time step is set to $T = 0.1$ s, and each episode is comprised of $100$ time steps with a duration of $10$ s. The coefficients in the reward function of ~\eqref{eq13} are set to $\alpha = \beta = 0.1$.

The FH-DDPG algorithm \cite{Lei2020} is adopted to solve the platoon
control problems, which adapts the DDPG algorithm for improving the
overall performance and convergence of finite-horizion
problems. Specifically, the DDPG algorithm is embedded
  into a finite-horizon value iteration framework. A pair of actor and
  critic networks are trained for each time step by backward
  induction, i.e., the agent starts from training the actor and critic
  networks of the last time step, and propagates backward in time
  until the networks of the first time step are trained. In training
  for each time step, the DDPG algorithm is used to solve a one-period
  MDP where the target networks are fixed to be the trained actor and
  critic networks of the next time step. For the detailed pseudocode
  of FH-DDPG, please refer to \cite{Lei2020}. The hyper-parameters
used for training are summarized in Table~\ref{table_2}, the values of
which were selected by performing a grid search as in
\cite{Mnih2015}. The sizes of the neural networks in the simulation
are given in Table~\ref{table_2}. There are three hidden layers in the
actor and critic networks, where the number of neurons in each layer
is $400$, $300$, and $100$, respectively. Note that the size of the
input layer for the actor is decided by the state dimension of
different SSDPs. For the critic network, an additional $1$-dimensional
action input is fed to the second hidden layer. The size of replay
buffer and batch are set to be $20,000$ and $128$ in all the
experiments, respectively. When the replay buffer is full, the oldest
sample will be discarded before a new sample is stored into the
buffer.

\begin{table}[htb!]
	\centering
	\caption{Technical constraints and operational parameters of the platoon control environment}
	\begin{tabular}{lr}
		\hline
		\textbf{Parameter} & \textbf{Value} \\
		\hline
		Interval for each time step $T$ & $0.1$ s\\
		\hline
		Total time steps in each episode $K$ &$100$\\
		\hline
	    Time constant for leader $0$ $\tau_0$ & $0.45$ s\\
		\hline
		Time gap $h_i$ & $0.3$ s\\
		\hline
		Max control input $u_{\rm max}$ & $2.6\rm{m/s^2}$\\
		\hline
		Min control input $u_{\rm min}$ & $-2.6\rm{m/s^2}$\\
		\hline
		Reward coefficients $\alpha, \beta$ & $\alpha = \beta = 0.1$\\
		\hline
	\end{tabular}
	\label{table_1}
\end{table}

\begin{table}[htb!]
	\centering
	\caption{Hyperparameters of the FH-DDPG algorithm}
	\begin{tabular}{ lr }
		\cline{1-2}
		\textbf{Parameter} & \textbf{Value} \\
		\cline{1-2}
		Actor network size & 400, 300, 100 \\ 
		\cline{1-2}
		Critic network size & 400, 300, 100 \\ 
		\cline{1-2}
		Actor activation function & relu, relu, relu, tanh \\ 
		\cline{1-2}
		Critic activation function & relu, relu, relu, linear \\ 
		\cline{1-2}
		Actor learning rate&1e-5 \\
		\cline{1-2}
		Critic learning rate &1e-4 \\
		\cline{1-2}
		Replay buffer size & 20000\\
		\cline{1-2}
		Batch size &128 \\
		\cline{1-2}
		Reward scale & 5e-3\\
		\cline{1-2}
		Noise type &\makecell[r]{Ornstein-Uhlenbeck Process with \\$\theta = 0.15$ and $\sigma = 0.5$} \\
\cline{1-2}
\makecell[l]{weights/\\ biases initialization} &\makecell[r]{Random uniform distribution \\$[-3\times 10^{-3},3\times 10^{-3}]$ (final layer) \\ $[-\frac{1}{\sqrt{fan-in}},\frac{1}{\sqrt{fan-in}}]$ (other layers)}\\

\cline{1-2}
	\end{tabular}
	\label{table_2}
\end{table}

\subsection{Training and testing results of two-vehicle scenario}
We perform simulations for SSDPs $\mathrm{P1}$, $\mathrm{P2}$, and $\mathrm{P3}$ under the two-vehicle scenario. The time constant $\tau_i$ for the follower is set to $0.5$s. We set the initial state to $S_{i,1}^{(\mathrm{P1})} = [2.5, 2.5, 0]^{\mathrm{T}}$. The control input $u_{i-1}$ of the predecessor is set to a sequence of independent random variables having Gaussian distribution.
\subsubsection{Performance across 5 runs}
The individual, average, and best observed performance as well as the standard errors across 5 runs are reported in Table~\ref{table_3} for $\mathrm{P1}$, $\mathrm{P2}$, and $\mathrm{P3}$. For each run, the individual performance is obtained by averaging the returns (cumulative rewards per episode) over $200$ test episodes after training is completed. We can observe that for each run, the individual performance of $\mathrm{P2}$ is always higher than that of $\mathrm{P1}$, which is consistent with Lemma 1. Moreover, $\mathrm{P3}$ shows the best performance among the three SSDPs, which agrees with Lemma 2. As shown in Table~\ref{table_3}, the standard error of $\mathrm{P3}$ is lower than those of $\mathrm{P1}$ and $\mathrm{P2}$, which indicates that the performance of $\mathrm{P3}$ is more stable than that of the other two problems.

\begin{table*}[htb!]
	\renewcommand{\arraystretch}{1}
	\setlength{\extrarowheight}{1pt}
	\centering
	\caption{Performance after training across $5$ different runs. Each run has $100$ time steps in total. We report the individual, average, best observed performance and standard errors (across 5 runs) for SSDPs $\mathrm{P1}$, $\mathrm{P2}$, and $\mathrm{P3}$ with FH-DDPG.}
	\begin{tabular}{ c|cccccccc}
		\hline
		\multirow{2}{*}{\textbf{SSDPs}}&\multicolumn{8}{c}{\textbf{Performance}} \\
		\cline{2-9}
		&\textbf{Run 1}&\textbf{Run 2}&\textbf{Run 3}&\textbf{Run 4}&\textbf{Run 5}&\textbf{Max}&\textbf{Average}&\textbf{Std Error}\\
		\hline
		$\mathrm{P1}$&-1.9263&-1.9104&-1.9870&-1.9310&-1.8949&-1.8949&-1.9299&0.0349\\ 
		\cline{1-9}
	$\mathrm{P2}$ &-1.9067&-1.8806&-1.9222&-1.8885&-1.9401 &-1.8806&-1.9076&0.0243 \\ 
		\cline{1-9}
		$\mathrm{P3}$&-1.8971&-1.8730&-1.8964&-1.8794&-1.8751&-1.8730&-1.8842&0.0105 \\
		\hline	
	\end{tabular}
\label{table_3}
\end{table*}

\subsubsection{Convergence properties}
\begin{figure}[htb!]
	\centering
	\includegraphics[scale=0.8]{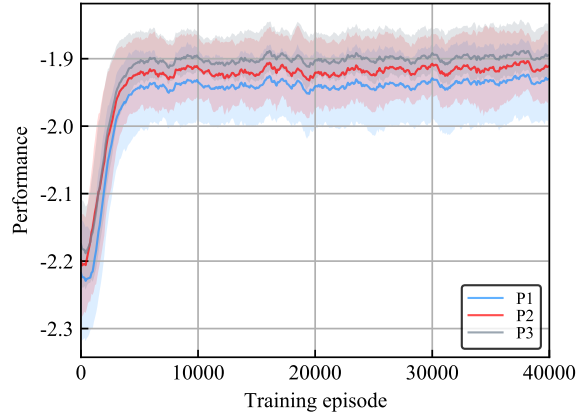}
	\caption{Average performance across 5 runs for SSDPs $\mathrm{P1}$, $\mathrm{P2}$, and $\mathrm{P3}$ with FH-DDPG. The vertical axis corresponds to the average performance across 5 runs and the shaded areas indicate the standard errors of three SSDPs.}
	\label{fig_4}
\end{figure}

The performance of control policies is evaluated periodically during
training by testing them without exploration noise. 
  Specifically, we run $10$ test episodes after every $100$ training
  episodes, and calculate the average cumulative rewards over the $10$
  test episodes as the performance for the latest $100$ training
  episode. The performance as a function of the number of training
episodes for $\mathrm{P1}$, $\mathrm{P2}$ and $\mathrm{P3}$ is given
in Fig.~\ref{fig_4}, where the curves correspond to the average
performance across 5 runs and the shaded areas indicate the standard
errors. Fig.~\ref{fig_4} shows that the convergence rate of the three
SSDPs is similar. Moreover, it can be observed that the shaded areas
of $\mathrm{P3}$ is much smaller than those of $\mathrm{P1}$ and
$\mathrm{P2}$, which indicates that $\mathrm{P3}$ performs more stably
across different runs than the other two SSDPs.


\subsubsection{Accuracy of Q-value estimations}
\begin{figure*}[t!]
	\centering
	\subfloat[$\mathrm{P1}$.]{\includegraphics[scale=0.31]{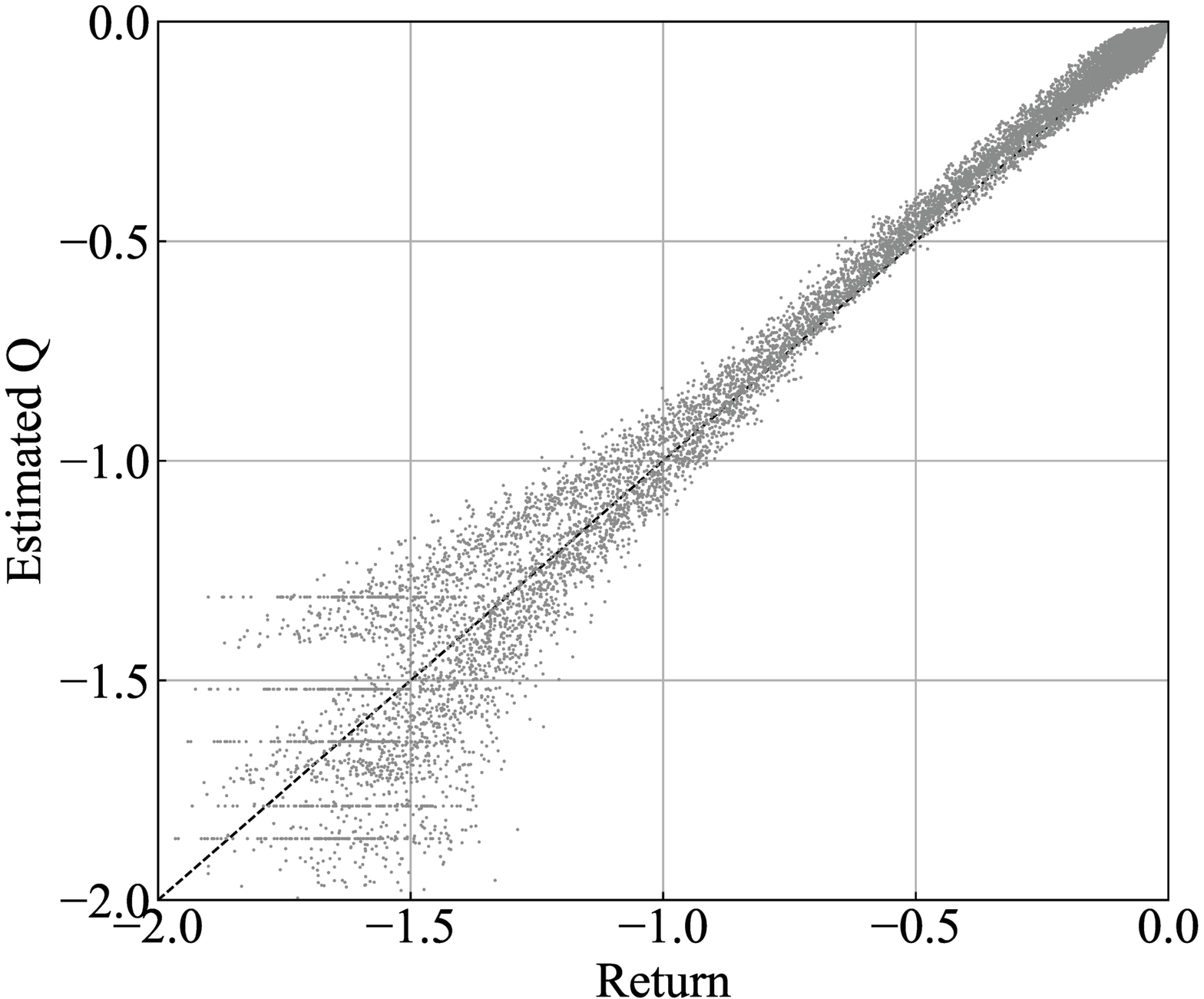}\label{fig_sub1}}
	\hfil
	\subfloat[$\mathrm{P2}$.]{\includegraphics[scale=0.31]{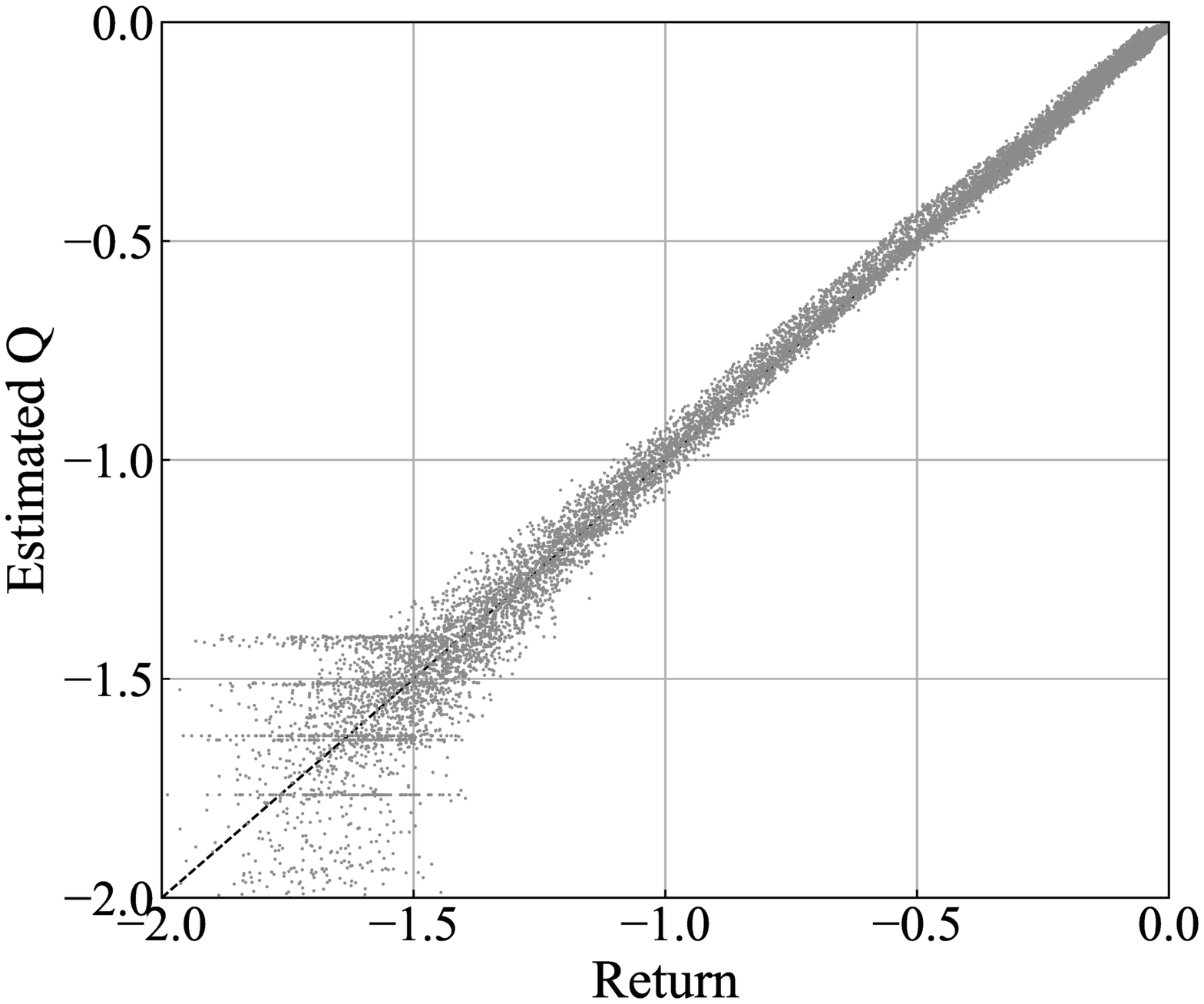}\label{fig_sub2}}
	\hfil
	\subfloat[$\mathrm{P3}$.]{\includegraphics[scale=0.31]{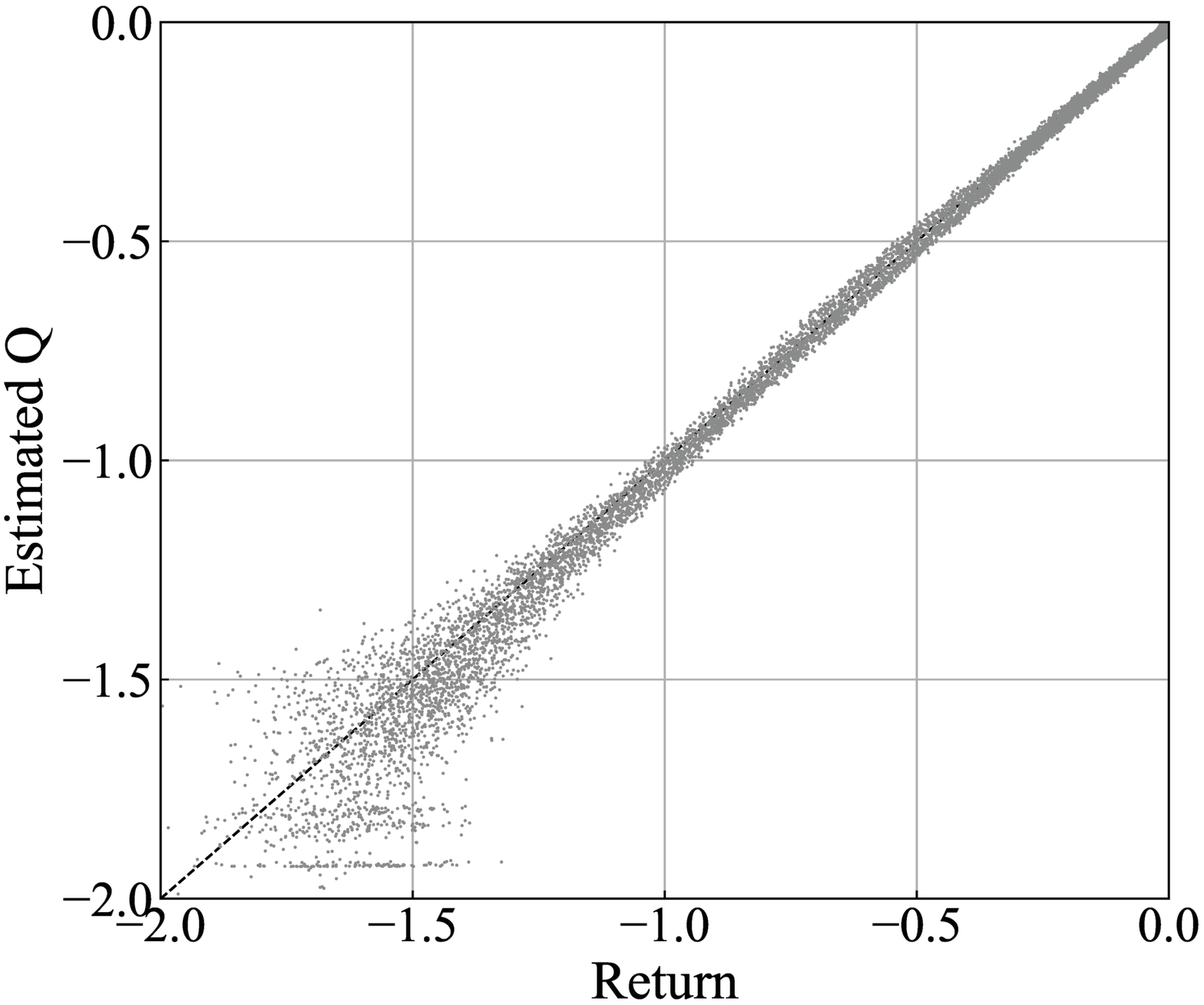}\label{fig_sub3}}
	\caption{Scatter plot showing estimated Q-values versus observed returns from test episodes on 5 runs. The vertical axis corresponds to the estimated Q-values while the horizontal axis corresponds to the true Q-values.}
	\label{fig_5}
\end{figure*}

As learning accurate Q-values is very important for the success of actor-critic algorithms, we examined the Q-values estimated by the critic after training by comparing them to the true returns seen on the test episodes. Fig.~\ref{fig_5} shows that compared to $\mathrm{P2}$ and $\mathrm{P3}$, the estimated Q-values of $\mathrm{P1}$ are more scattered and deviate farther from the true returns, especially at the beginning of an episode when the Q-values are more negative. Given the better accuracy of the estimated Q-values, $\mathrm{P2}$ and $\mathrm{P3}$ are able to learn better policies compared to $\mathrm{P1}$, as shown in Table \ref{table_3}. 

Moreover, the inaccuracy in estimated Q-values also explains why the ranking of estimated Q-values for the three problems in Fig. \ref{fig_4} is inconsistent with the performance ranking in Table \ref{table_3}.

\subsubsection{Test results for one episode}
\begin{figure}[htb!]
	\centering
	\includegraphics[scale=0.38]{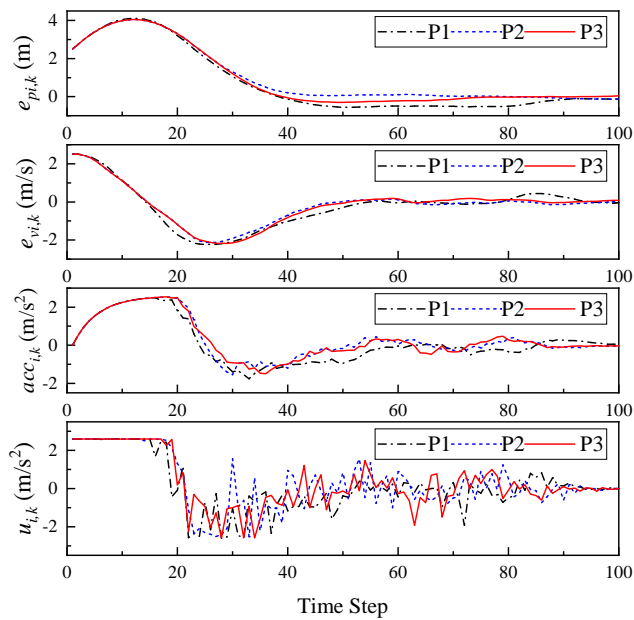}
	\caption{Results for a specific $10$s test episode under the two-vehicle scenario. The driving status $e_{pi,k}$, $e_{vi,k}$, $acc_{i,k}$ and control input $u_{i,k}$ of $\mathrm{P1}$, $\mathrm{P2}$, and $\mathrm{P3}$ are represented as different curves, respectively.}
	\label{fig_6}
\end{figure}

Here we focus our attention on a specific test episode having $100$ time steps, and plot the control input $u_{i,k}$ along with the driving status $e_{pi,k}$, $e_{vi,k}$, and $acc_{i,k}$ for all the time steps $k\in\{0,1,\cdots,99\}$. Fig.~\ref{fig_6} shows the results for $\mathrm{P1}$, $\mathrm{P2}$, and $\mathrm{P3}$, where it can be observed that the overall shapes of the curves for the three problems look very similar. At the beginning of the episode, namely for time steps $k\leq20$, the control input $u_{i,k}$ remains the maximum value $u_{\rm max} = 2.6\rm{m/s^2}$ to increase the acceleration $acc_{i,k}$ as promptly as possible, so that the control errors $e_{pi,k}$ and $e_{vi,k}$ can be promptly reduced. Since the initial velocity error $e_{{vi},1}=2.5$ is a positive value, $e_{pi,k}$ increases first for $k<10$ and then decreases, when $e_{vi,k}$ becomes negative. At around $k=40$, the control input $u_{i,k}$ and driving status $e_{pi,k}$, $e_{vi,k}$, $acc_{i,k}$ become approximately $0$. Beyond that, the values fluctuate around $0$, with $u_{i,k}$ trying to track the random control input $u_{i-1,k}$ of the predecessor.\par 

A closer examination of Fig.~\ref{fig_6} shows that the control error $e_{pi,k}$ and $e_{vi,k}$ for $\mathrm{P1}$ exhibits higher fluctuation than those of $\mathrm{P2}$ and $\mathrm{P3}$. For example, $e_{pi,k}$ decreases to about $-1$ m at around $k = 50$. The curves of $e_{pi,k}$ and $e_{vi,k}$ for $\mathrm{P2}$ and $\mathrm{P3}$ are relatively close. However, the control input $u_{i,k}$ of $\mathrm{P3}$ has lower fluctuation than those of $\mathrm{P2}$, which means that the vehicle supported by $\mathrm{P3}$ drives more smoothly. Note that $e_{pi,k}$, $e_{vi,k}$, and $u_{i,k}$ affect the reward function as defined in \eqref{eq13}, so the results in Fig.~\ref{fig_6} further validate the performance ranking in Table \ref{table_3}.\par

\subsection{Training and testing results of platooning scenario}
We perform simulations for the platooning scenario with $5$ vehicles excluding the leader (i.e., vehicle $0$). We consider that the control polices of all the followers are trained under SSDP P4, except for the second last but one vehicle (i.e., vehicle $4$), which is trained under $\mathrm{P4}$, $\mathrm{P5}$, $\mathrm{P6}$ as well as the other SSDPs given in Table \ref{table_4}. Moreover, we simulate a heterogeneous platoon where the time constants $\tau_{i}$ are given in Table \ref{table_5} for the vehicles $i\in\{1,2,3,4,5\}$. By comparing the performance of vehicle $4$ under different SSDPs, we can gain useful insights into the impact of V2X information on DRL-based platoon control.\par

\begin{table}[htb!]
	\centering
	\caption{Time constants of vehicles in the platoon}
	\begin{tabular}{cccccc}
		\hline
		Vehicle index $i$ & $1$ & $2$ & $3$ & $4$ & $5$\\
		\hline
		Time constant $\tau_i$ & $0.5$ & $0.25$ & $0.2$ & $0.1$ & $0.3$ \\ 
			\hline
	\end{tabular}
	\label{table_5}
\end{table}

We set the initial state for each of the $5$ following vehicles to be $S_{i,1}^{(\mathrm{P1})} = [1.5, -1, 0]^{\mathrm{T}}$, $\forall i\in\{1,2,3,4,5\}$. The control input $u_{0}$ of the leader is set to a sequence of independent random variables obeying the Gaussian distribution. \par

Similar to the two-vehicle scenario, the individual, average, and best observed performance as well as the standard errors across $5$ runs are reported in Table~\ref{table_6}. For each run, the individual performance is obtained by averaging the returns (cumulative rewards per episode) over $200$ test episodes after training is completed. \par

\begin{table*}[htb!]
	\renewcommand{\arraystretch}{1.2}
	\setlength{\extrarowheight}{1pt}
	\centering
	\caption{Performance after training across 5 different runs. Each run has $100$ time steps in total. We report the individual, average, best observed performance and standard errors (across 5 runs) for different SSDPs in platoon scenario with FH-DDPG.}
	\begin{tabular}{ c|cccccccc}
		\hline
		\multirow{2}{*}{\textbf{Problem}}&\multicolumn{8}{c}{\textbf{Performance}} \\
		\cline{2-9}
		&\textbf{Run 1}&\textbf{Run 2}&\textbf{Run 3}&\textbf{Run 4}&\textbf{Run 5}&\textbf{Max}&\textbf{Average}&\textbf{Std Error}\\
		\hline
				$\mathbf{P4}$&-0.1177 &-0.1115 &-0.1111 &-0.1231 &-0.1262 &-0.1111 &-0.1179 &0.0068  \\ 
		\cline{1-9}
		$\mathbf{P_{PF2}}$ &-0.1135&-0.1159 &-0.1192 &-0.1307 &-0.1129&-0.1129 &-0.1184&0.0073 \\
		\cline{1-9}
		$\mathbf{P_{PLF}}$ &-0.1172 &-0.1131 &-0.1179 &-0.1175 &-0.1104 &-0.1104 &-0.1152 &0.0033  \\
		\cline{1-9}
		$\mathbf{P_{TPF}}$ &-0.1071 &-0.1080 &-0.1116 &-0.1102 &-0.1068&-0.1068 &-0.1087 & 0.0021 \\
		\cline{1-9} 
		$\mathbf{P_{TPLF}}$ &-0.1110 &-0.1101 &-0.1123 &-0.1172 &-0.1090 &-0.1090 &-0.1119 &0.0032\\
		\cline{1-9} 
		$\mathbf{P5}$ &-0.1060 &-0.1024&-0.1032 &-0.1064&-0.1022 &-0.1022 &-0.1040 &0.0020  \\ 
		\cline{1-9}
		$\mathbf{P6}$ &-0.1086 &-0.1048 &-0.1065 &-0.1106&-0.1104 &-0.1048 &-0.1082 &0.0025 \\
		\hline	                                                                                
	\end{tabular}                                                                           
	\label{table_6}
\end{table*}

We first compare the performance of $\mathrm{P4}$, $\mathrm{P5}$, and $\mathrm{P6}$. Observe from Table~\ref{table_6} that $\mathrm{P5}$ performs consistently better than $\mathrm{P4}$ in each individual run. Moreover, the standard error of $\mathrm{P5}$ is also lower than that of $\mathrm{P4}$, showing that the performance of $\mathrm{P5}$ is more stable than $\mathrm{P4}$. The observations agree with Lemma 3, stating that the optimal policy of $\mathrm{P5}$ is at least as good as that of $\mathrm{P4}$. Moreover, it can be deduced from the results that the gain due to the availability of V2X information from all the preceding vehicles (i.e., vehicle $1$, $2$, $3$) offsets the loss due to the function approximation error resulting from having a higher state dimension. \par  

Meanwhile, Table~\ref{table_6} shows that $\mathrm{P5}$ also performs consistently better than $\mathrm{P6}$ in terms of all the performance metrics. According to Lemma 4, the optimal policies of $\mathrm{P5}$ and $\mathrm{P6}$ have the same performance. However, as $\mathrm{P6}$ has larger state space than $\mathrm{P5}$, the DRL policy of $\mathrm{P6}$ performs worse than that of $\mathrm{P5}$ due to the larger function approximation error in actor and critic networks.  \par  

Although $\mathrm{P5}$ performs better than $\mathrm{P4}$, it requires that both the driving status and control inputs be transmitted from all the preceding vehicles, which involves high communication overhead. Now we compare the performance of the other SSDPs in Table \ref{table_4} to see if we can reduce the communication overhead while still achieving a relative good performance. Observe from Table~\ref{table_6} that the rankings in terms of the individual performance vary slightly across different runs. The ranking in terms of the average performance for the different SSDPs is $\mathrm{P5}>\mathrm{P_{TPF}}>\mathrm{P_{TPLF}}>\mathrm{P_{PLF}}>\mathrm{P4}>\mathrm{P_{PF2}}$. This ranking is consistent with the rankings in terms of the maximum performance and standard error. \par

In order to gain further insights into the performance ranking in Table~\ref{table_6}, we evaluate the value of V2X information for DRL-based platoon control by using \eqref{eq38} to derive the conditional KL divergence of $T^{S_{i}^{(\mathrm{Pm})}}\otimes T^{S_{i,k}^{(\mathrm{Pn})}\backslash S_{i,k}^{(\mathrm{Pm})}}$ from $T^{S_{i}^{(\mathrm{Pn})}}$. We fix $\mathrm{Pn}=\mathrm{P5}$ and let $\mathrm{Pm}$ be any other SSDP in Table \ref{table_4}. In other words, we evaluate "\emph{How much better would we be able to predict the future state if we included
the additional V2X information in $\mathrm{P5}$ as compared to $\mathrm{Pm}$, versus we
didn't?}". Lower KL divergence indicates less value for the additional information in $\mathrm{P5}$, and higher chance that $\mathrm{Pm}$ can achieve similar performance to $\mathrm{P5}$, despite its lower communication overhead. \par 

In order to obtain both the joint and conditional probability distributions on the R.H.S of \eqref{eq38}, we perform Monte-Carlo simulation for $200$ test episodes with the trained actors of vehicles $1$, $2$, and $3$, and keep a record of all the states and actions $\{S_{j,k}^{(\mathrm{P1})},u_{j,k}\}_{j=0}^{3}$. Then for each time step $k\in\{1,\cdots,K-1\}$, a quantization process is applied to the continuous states and actions $\{S_{j,k}^{(\mathrm{P1})},u_{j,k}\}_{j=0}^{3}$ as well as $u_{3,k+1}$ to derive the probability distributions in \eqref{eq38}, which are used for determining the KL divergence at that time step.\par 

\begin{figure}[htb!]
	\centering
	\includegraphics[scale=0.28]{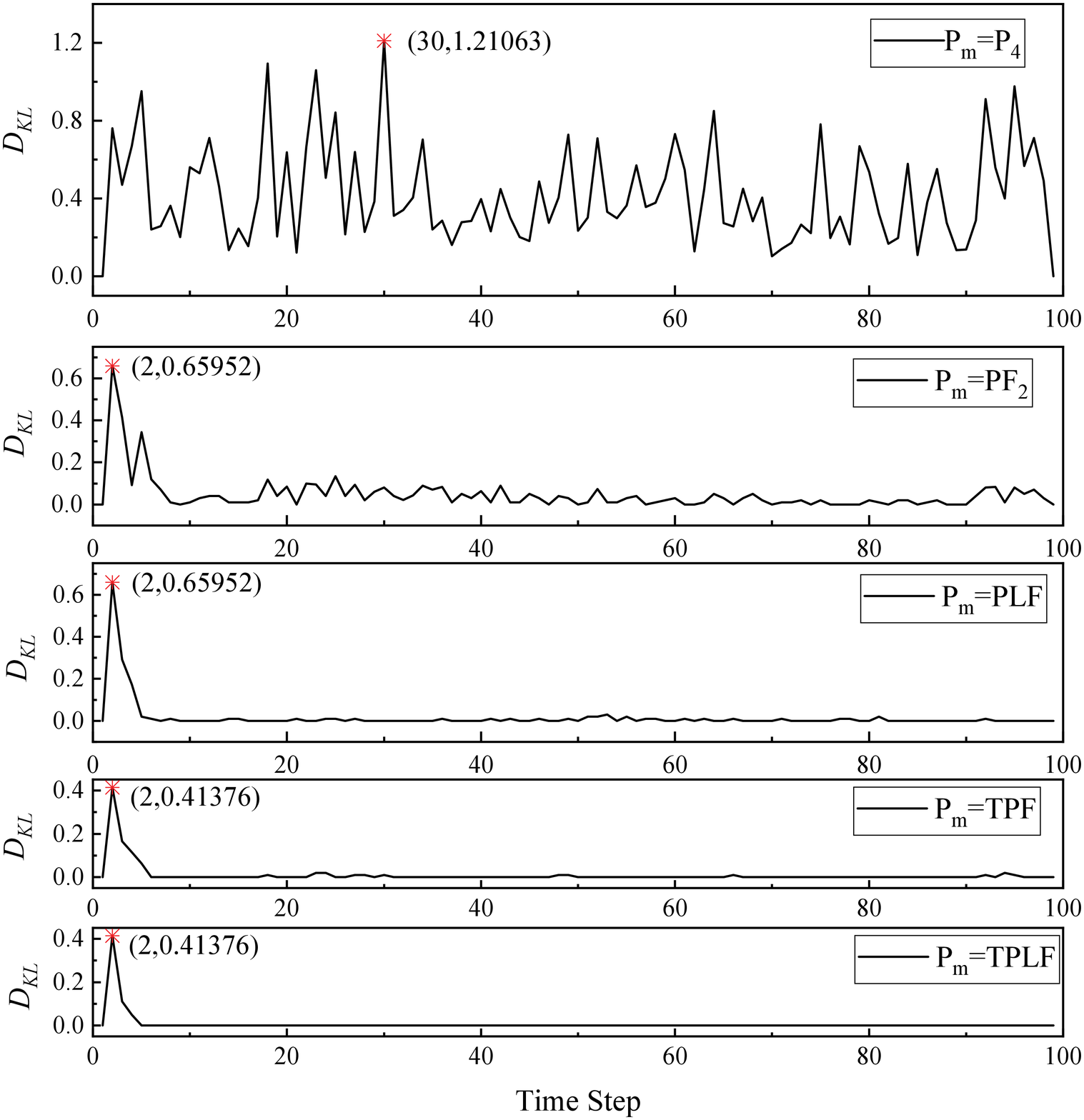}
	\caption{The conditional KL divergence $D_{KL}(T^{S_{i}^{(\mathrm{Pn})}}||T^{S_{i}^{(\mathrm{Pm})}}\otimes T^{S_{i,k}^{(\mathrm{Pn})}\backslash S_{i,k}^{(\mathrm{Pm})}})$ for each time step when $\mathrm{Pn=P5}$.}
	\label{fig_7}
\end{figure}

Figure \ref{fig_7} shows the KL divergence for each time step. It can be seen that when $\mathrm{Pm=P4}$, the KL divergences are relatively high for every time step. Meanwhile for all the other SSDPs, the KL divergences are relatively high for the first few time steps, but decays for the rest of the time steps. This shows that firstly, the value of the additional information in $\mathrm{P5}$ on platoon control as compared to the other SSDPs is high for the first few time steps. Secondly, compared to $\mathrm{P4}$, the other SSDPs have lower KL divergence, and thus transmitting the corresponding V2X information and including them in the state space can help better predict the next state in $\mathrm{P4}$.   \par

Now we examine the relationship between the KL divergence of an SSDP in Figure \ref{fig_7} and its DRL-based control performance in Table \ref{table_6}. Note that except for $\mathrm{P4}$, $\mathrm{P_{PF2}}$ has the highest KL divergence, as shown in Figure \ref{fig_7}. Its peak KL divergence is approximately the same as that of the $\mathrm{P_{PLF}}$, with a value of $0.66$ at time step $2$, but its KL divergence for later time steps is higher than those of $\mathrm{P_{PLF}}$ and other SSDPs (except for $\mathrm{P4}$). As analyzed earlier, we can see from Table \ref{table_6} that $\mathrm{P_{PF2}}$ performs the worst, even worse than $\mathrm{P4}$ in terms of both its average performance and standard error, although its maximum performance is better than that of $\mathrm{P4}$. This shows that the performance gain of including $e_{p3,k}$ and $e_{v3,k}$ in the state variable cannot offset the loss due to the increased state dimension on average. \par

Meanwhile, it can be observed from Figure \ref{fig_7} that the KL divergence of $\mathrm{P_{TPF}}$ is lower than that of $\mathrm{P_{PLF}}$. The former has a peak value of $0.41$ at time step $2$, while the latter has a peak value of $0.66$. This shows that the V2X information concerning the second immediate vehicle, i.e., $S_{2,k}^{(\mathrm{P1})}$ and $u_{2,k}$, has higher value than the V2X information on the leader, i.e., $acc_{0,k}$ and $u_{0,k}$. Moreover, it can be seen from Figure \ref{fig_7} that the KL divergences of $\mathrm{P_{TPF}}$ and $\mathrm{P_{TPLF}}$ are very similar, both of which have a peak value of $0.42$ at time step $2$, with the value of $\mathrm{P_{TPLF}}$ slightly smaller than that of $\mathrm{P_{TPF}}$ at later time steps. This shows that if the V2X information on the second immediate vehicle is available, further information on the leader will have little value in helping to predict the future states.\par 

The above insights on the KL divergence agree with the ranking of SSDPs seen in Table \ref{table_6}. Note that the performance of $\mathrm{P_{TPF}}$ is only slightly worse than that of $\mathrm{P5}$, while better than those of the other SSDPs both in terms of the average and maximum performance. On the other hand, $\mathrm{P_{TPF}}$ only requires V2X information from the second immediate vehicle instead of all the preceding vehicles as in $\mathrm{P5}$. Therefore, if the communication resources are limited, $\mathrm{P_{TPF}}$ is a good alternative for $\mathrm{P5}$. Another interesting observation is that the performance of $\mathrm{P_{TPLF}}$ is worse than that of $\mathrm{P_{TPF}}$ in terms of both the average and maximum performance, although its standard error is slightly lower than that of $\mathrm{P_{TPF}}$. This shows that compared to $\mathrm{P_{TPF}}$, the modest performance gain due to the availability of leader information cannot offset the loss due to having a higher state dimension in $\mathrm{P_{TPLF}}$, although $\mathrm{P_{TPLF}}$ performs a little more stably than $\mathrm{P_{TPF}}$. \par

\section{Conclusion}
In this paper, we have formalized the platoon control problems associated with different IFT and V2X information into different SSDP models, and provided theorems and lemmas for comparing the performance of their optimal policies. It has been shown that when there is only a single following vehicle, transmission of the acceleration and control input from the preceding vehicle can help improve the optimal control performance. When there are multiple following vehicles in a platoon, and the objective of each vehicle is to optimize its own performance, information transmission from all the preceding vehicles instead of only the immediate preceding vehicle could help improve the optimal policy, while information transmission from the following vehicles does not help.\par 

Moreover, we have used the conditional KL divergence for quantifying the value of V2X information in DRL-based control policies for the SSDPs. V2X information associated with larger values can help to better improve the DRL-based platoon control performance, and thus should be given higher priority in transmission, especially when the communication resources are limited.\par 

We have performed simulations for verifying our analytical results. For a platoon with $5$ following vehicles, our simulation results have shown that including V2X information from all the preceding vehicles achieved the best DRL-based control performance, while including V2X information from only the immediate and second immediate preceding vehicles struck a compelling trade-off between the control performance and communication overhead.\par

In this paper, we have focused our attention on decentralized platoon
control, where each vehicle optimizes its own performance. When the
objective is to optimize the global performance (i.e., sum of local
performances), the SSDPs become multi-agent problems and we will
explore the value of V2X information in this multi-agent setting in
our future work. Moreover, we will also consider the
  impact of the actuator delay and communications delay on the value
  of V2X information in the future.\par

\appendix
\subsection{Proof of Theorem 1}
Note that the original SSDP is not an MDP according to Remark 1, as the exogenous information transition function $f^{W}(S_{k},W_{k},\xi_{k})$ depends on $S_{k}$ or $W_{k}$ or both. On the other hand, for the augmented-state SSDP, the system state transition function becomes
	\begin{align}
\label{aeq3}
\tilde{S}_{k+1}&=
\begin{pmatrix}
S_{k+1}\\ W_{k+1}
\end{pmatrix}
=
\begin{pmatrix}
f^{S}(S_{k},a_{k},W_{k})\\ 
f^{W}(S_{k},W_{k},\xi_{k})
\end{pmatrix} =f^{\tilde{S}}(\tilde{S}_{k},a_{k},\tilde{W}_{k}),
\end{align}
where the exogenous information $\tilde{W}_{k}=\xi_{k}$ is an independent random variable with given distribution. Therefore, the augmented-state SSDP becomes an MDP. It is straightforward to see that the optimal policy for the augmented-state SSDP $\tilde{\pi}^{*}(\tilde{S}_{k})$ could improve over that of the original problem $\pi^{*}(S_{k})$ as the former policy is based on an MDP while the later is based on a non-Markovian SSDP. In other words, the original SSDP only has partial observability while the augmented-state SSDP has full observability.\par

\subsection{Proof of Theorem 2}
Note that the original SSDP is an MDP, while the augmented-state SSDP is also an MDP having a system state transition function
\begin{align}
\label{aeq7}
\tilde{S}_{k+1}&=
\begin{pmatrix}
S_{k+1}\\ W_{k+1}
\end{pmatrix}
=
\begin{pmatrix}
f^{S}(S_{k},a_{k},W_{k})\\ 
f^{W}(f^{S}(S_{k},a_{k},W_{k}),\xi_{k})
\end{pmatrix}\IEEEnonumber \\	
&=f^{\tilde{S}}(\tilde{S}_{k},a_{k},\tilde{W}_{k}),
\end{align}
where the exogenous information $\tilde{W}_{k}=\xi_{k}$ is an independent random variable with given distribution.\par

Let $V_{k}^{*}(S_{k})$ and $\tilde{V}_{k}^{*}(\tilde{S}_{k})$ denote the value functions under the optimal policies $\pi^{*}$ and $\tilde{\pi}^{*}$ for the original SSDP and augmented-state SSDP, respectively. Define $\tilde{V}_{k}^{*}(S_{k})=\mathrm{E}_{W_{k}}[\tilde{V}_{k}^{*}(\tilde{S}_{k})|S_{k}]$.\par 

Note that $\tilde{J}^{*}=\mathrm{E}_{\tilde{S}_{1}}[\tilde{V}_{1}^{*}(\tilde{S}_{1})]=\mathrm{E}_{S_{1}}[\tilde{V}_{1}^{*}(S_{1})]$ and $J^{*}=\mathrm{E}_{S_{1}}[V_{1}^{*}(S_{1})]$. Therefore, in order to prove that $\tilde{J}^{*}\geq J_{i}^{*}$, it is sufficient to prove 
\begin{equation}
\label{aeq6}
\tilde{V}_{k}^{*}(S_{k})\geq V_{k}^{*}(S_{k}), \forall \ S_{k} \ \mathrm{and} \ k.
\end{equation} 

We will show \eqref{aeq6} by induction. For the last time step $K$, we have
\begin{equation}
\label{aeq1}
\tilde{V}_{K}^{*}(S_{K})= \mathrm{E}_{W_{K}}\Big[\max_{\tilde{\mu}_{K}(\tilde{S}_{K})}R\big(\tilde{S}_{K},\tilde{\mu}_{K}(\tilde{S}_{K})\big)|S_{K}\Big], \forall \ S_{K},
\end{equation}
\begin{equation}
\label{aeq2}
V_{K}^{*}(S_{K})= \max_{\mu_{K}(S_{K})}\mathrm{E}_{W_{K}}\Big[R\big(\tilde{S}_{K},\mu_{K}(S_{K})\big)|S_{K}\Big], \forall \ S_{K}.
\end{equation}
According to \eqref{aeq1} and \eqref{aeq2}, we have
\begin{equation}
\label{aeq10}
\tilde{V}_{K}^{*}(S_{K})\geq V_{K}^{*}(S_{K}), \forall \ S_{K},
\end{equation} 
\noindent(since we generally have $\mathrm{E}[\max\{\cdot\}]\geq\max\{E[\cdot]\}$ according to Jensen's inequality). Therefore, the optimal action $\tilde{\mu}_{K}(\tilde{S}_{K})$ for the augmented-state SSDP problem is at least as good as that for the original SSDP $\mu_{K}(S_{K})$ at time step $K$. 

Assume that
\begin{equation}
\label{aeq9}
\tilde{V}_{k+1}^{*}(S_{k+1})\geq V_{k+1}^{*}(S_{k+1}), \forall \ S_{k+1}.
\end{equation}

Consider the Bellman Equation for the original SSDP as 
\begin{align}
\label{aeq8}
&V_{k}^{*}(S_{k})=\max_{\mu_{k}(S_{k})}\Biggl\{\mathrm{E}_{W_{k}}\biggl[R\bigl(S_{k},\mu_{k}(S_{k}),W_{k}\bigr)+V_{k+1}^{*}\bigl(f^{S}(S_{k},\IEEEnonumber \\&\mu_{k}(S_{k}),W_{k})\bigr)\Big|S_{k}\biggr]\Biggr\},
\end{align}

\noindent and consider the Bellman Equation for the augmented-state problem as 
\begin{align}
\label{aeq4}
&\tilde{V}_{k}^{*}(\tilde{S}_{k}) =\max_{\tilde{\mu}_{k}(\tilde{S}_{k})}\Biggl\{R\bigl(\tilde{S}_{k},\tilde{\mu}_{k}(\tilde{S}_{k})\bigr)+\mathrm{E}_{\tilde{S}_{k+1}}\biggl[\tilde{V}_{k+1}^{*}(S_{k+1},\IEEEnonumber \\&W_{k+1}) \Big|\tilde{S}_{k}\biggr]\Biggr\}. 
\end{align}


Taking the expectation over $W_{k}$ conditioned on $S_{k}$ at both sides of \eqref{aeq4}, we have the following Bellman equation
\begin{align}
\label{aeq5}
&\tilde{V}_{k}^{*}(S_{k})=\mathrm{E}_{W_{k}}\bigl[\tilde{V}_{k}^{*}(\tilde{S}_{k})|S_{k}\bigr]\IEEEnonumber \\
&=\mathrm{E}_{W_{k}}\Biggl[\max_{\tilde{\mu}_{k}(\tilde{S}_{k})}\Biggl\{R\bigl(\tilde{S}_{k},\tilde{\mu}_{k}(\tilde{S}_{k})\bigr)+\mathrm{E}_{\tilde{S}_{k+1}}\biggl[\tilde{V}_{k+1}^{*}\Bigl(S_{k+1},W_{k+1}\Bigr) \IEEEnonumber \\
&\Big|\tilde{S}_{k}\biggr]\bigg|S_{k}\Biggr\}\Biggr] \IEEEnonumber \\
&\stackrel{(a)}{\geq}\max_{\mu_{k}(S_{k})}\biggl\{\mathrm{E}_{W_{k}}\Bigl[R\bigl(\tilde{S}_{k},\mu_{k}(S_{k})\bigr)+\mathrm{E}_{\tilde{S}_{k+1}}\bigl[\tilde{V}_{k+1}^{*}(S_{k+1},W_{k+1}) \IEEEnonumber \\
&|\tilde{S}_{k}\bigr]\big|S_{k}\Bigr]\biggr\} \IEEEnonumber \\
&\stackrel{(b)}{=}\max_{\mu_{k}(S_{k})}\Biggl\{\mathrm{E}_{W_{k}}\biggl[R\bigl(\tilde{S}_{k},\mu_{k}(S_{k})\bigr)|S_{k}\biggr]+\mathrm{E}_{\tilde{S}_{k+1}}\biggl[\tilde{V}_{k+1}^{*}\bigl(S_{k+1} \IEEEnonumber \\
&,W_{k+1}\bigr)|S_{k}\biggr]\Biggr\},  \IEEEnonumber \\
&\stackrel{(c)}{=}\max_{\mu_{k}(S_{k})}\Biggl\{\mathrm{E}_{W_{k}}\biggl[R\bigl(\tilde{S}_{k},\mu_{k}(S_{k})\bigr)\Big|S_{k}\biggr]+\mathrm{E}_{S_{k+1}}\biggl[\mathrm{E}_{W_{k+1}}\Bigl[\tilde{V}_{k+1}^{*}\IEEEnonumber \\
&\bigl(S_{k+1},W_{k+1}\bigr)|S_{k+1}\Bigr]\Big|S_{k}\biggr]\Biggr\},  \IEEEnonumber \\
&\stackrel{(d)}{=}\max_{\mu_{k}(S_{k})}\Bigl\{\mathrm{E}_{W_{k}}\bigl[R\bigl(\tilde{S}_{k},\mu_{k}(S_{k})\bigr)+\tilde{V}_{k+1}^{*}(S_{k+1})|S_{k}\bigr]\Bigr\}\IEEEnonumber \\
&\stackrel{(e)}{\geq}\max_{\mu_{k}(S_{k})}\Bigl\{\mathrm{E}_{W_{k}}\bigl[R\bigl(\tilde{S}_{k},\mu_{k}(S_{k})\bigr)+V_{k+1}^{*}(S_{k+1})|S_{k}\bigr]\Bigr\}\IEEEnonumber \\
&\stackrel{(f)}{=}V_{k}^{*}(S_{k}),
\end{align}
\noindent where (a) follows by interchanging the expectation and maximization (since we generally have $\mathrm{E}[\max\{\cdot\}]\geq\max\{E[\cdot]\}$ according to Jensen's inequality); (b) is due to the properties of conditional expectations; (c) is due to the transition function $W_{k+1}=f^{W}(S_{k+1},\xi_{k})$ in Definition 1, which states that $W_{k+1}$ may be dependent on $S_{k+1}$, but independent of $S_{k}$; (d) is due to the definition of $\tilde{V}_{k+1}^{*}(S_{k+1})$, and the state transition function $S_{k+1}=f^{S}(S_{k},a_{k},W_{k})$ in Definition 1; (e) follows from \eqref{aeq9}; and (f) follows from the Bellman equation for the original SSDP as given in \eqref{aeq8}. Thus \eqref{aeq6} is proved for all $k$ and the desired results are shown.\par

\subsection{Proof of Theorem 3}
We will revisit the proof of Theorem 2 in Appendix B, substituting in the two conditions seen in Theorem 3. If $W_{k}$ does not affect the reward function as in Condition (2) of Theorem 3, we have $R(S_{k},a_{k},W_{k})=R(S_{k},a_{k})$, which means that the expectation operator can be eliminated in \eqref{aeq1} and \eqref{aeq2}. Therefore, we have $\tilde{V}_{K}^{*}(S_{K})= V_{K}^{*}(S_{K}), \forall \ S_{K}$ in \eqref{aeq10}. Now, assume that $\tilde{V}_{k+1}^{*}(S_{k+1})= V_{k+1}^{*}(S_{k+1}), \forall \ S_{k+1}$, then (g) in \eqref{aeq5} becomes an equality. Moreover, if $W_{k}$ does not affect both the state transition and reward function as stated in Theorem 3, the expectation operator over $W_{k}$ can be eliminated at both sides of (c) in \eqref{aeq5}, and the inequality in (c) becomes an equality. Therefore, we can prove that $\tilde{V}_{k}^{*}(S_{k})= V_{k}^{*}(S_{k}), \forall \ k, S_{k}$, and thus prove Theorem 3.  

\subsection{Proof of Lemma 1}
Since $S_{i,k}^{(\mathrm{P2})}=[(S_{i,k}^{(\mathrm{P1})})^{\mathrm{T}},W_{i,k}^{(\mathrm{P1})}]^{\mathrm{T}}$, we can consider $\mathrm{P1}$ as the original SSDP given in Definition 1 and $\mathrm{P2}$ as the augmented-state SSDP given in Definition 2. Moreover, the transition function of exogenous information in $\mathrm{P1}$ is given in \eqref{eq12} as $W_{i,k+1}^{(\mathrm{P1})}=f^{W^{(\mathrm{P1})}}(W_{i,k}^{(\mathrm{P1})},u_{i-1,k})$, where $u_{i-1,k}$ is an independent random variable with given distribution. Therefore, Lemma 1a follows from Theorem 1.\par

Since $S_{i,k}^{(\mathrm{P3})}=[(S_{i,k}^{(\mathrm{P2})})^{\mathrm{T}},W_{i,k}^{(\mathrm{P2})}]^{\mathrm{T}}$, we can consider $\mathrm{P2}$ as the original SSDP and $\mathrm{P3}$ as the augmented-state SSDP. Moreover, the exogenous information in $\mathrm{P2}$ $W_{i,k}^{(\mathrm{P2})}=u_{i-1,k}$ is an independent random variable with given distribution. Therefore, Lemma 1b follows from Theorem 2. 

\subsection{Proof of Lemma 2}
Since $S_{i,k}^{(\mathrm{P5})}=[(S_{i,k}^{(\mathrm{P4})})^{\mathrm{T}},(W_{i,k}^{(\mathrm{P4}\_1)})^{\mathrm{T}},(W_{i,k}^{(\mathrm{P4}\_2)})^{\mathrm{T}}]^{\mathrm{T}}$, we can consider $\mathrm{P4}$ as the original SSDP and $\mathrm{P5}$ as the augmented-state SSDP. Moreover, the transition function of exogenous information in $\mathrm{P4}$ is given in \eqref{eq32} as $W_{i,k+1}^{(\mathrm{P4\_12})}=f^{W^{(\mathrm{P4\_12})}}(W_{i,k}^{(\mathrm{P4\_12})},S_{i,k}^{(\mathrm{P4})})$. Therefore, Lemma 2 follows from Theorem 1. 

\subsection{Proof of Lemma 3}
Consider $\bar{W}_{i,k}^{(\mathrm{P5})}=[S_{i+1,k}^{(\mathrm{P1})},\cdots,S_{N-1,k}^{(\mathrm{P1})}]^{\mathrm{T}}$ as the exogenous information in $\mathrm{P5}$ in addition to $u_{0,k+1}$. Therefore, as $S_{i,k}^{(\mathrm{P6})}=[(S_{i,k}^{(\mathrm{P5})})^{\mathrm{T}},(\bar{W}_{i,k}^{(\mathrm{P5})})^{\mathrm{T}}]^{\mathrm{T}}$, we can consider $\mathrm{P5}$ as the original SSDP and $\mathrm{P6}$ as the augmented-state SSDP. It is plausible that $\bar{W}_{i,k}^{(\mathrm{P5})}$ will affect neither the reward function nor the system transition function in $\mathrm{P5}$, and thus according to Theorem 3, the augmented-state SSDP including $\bar{W}_{i,k}^{(\mathrm{P5})}$ as part of the state will result in an optimal policy that has the same performance as that in $\mathrm{P5}$.\par

\bibliography{platoon}{}
\bibliographystyle{IEEEtran}

\end{document}